\documentclass[12pt,preprint,tighten,iop, apj]{aastex}
\usepackage[english]{babel}
\usepackage[]{times, graphicx}
\newcommand{\pref}{\protect\ref}
\newcommand{\soho}{{\em SOHO{}}}
\newcommand{\sdo}{{\em SDO{}}}
\newcommand{\degree}{${^\circ{}}$}

\begin{document}

\shorttitle{Deciphering Solar Magnetic Activity \-- Paper 1}
\shortauthors{S.W. McIntosh et al.}

\title{Deciphering Solar Magnetic Activity I: On The Relationship Between The Sunspot Cycle And The Evolution Of Small Magnetic Features}



\author{Scott W. McIntosh\altaffilmark{1}, Xin Wang\altaffilmark{1,2}, Robert J. Leamon\altaffilmark{3}, \\
Rachel Howe\altaffilmark{4}, Larisza D. Krista\altaffilmark{5}, Anna V. Malanushenko\altaffilmark{6}, \\
Jonathan W. Cirtain\altaffilmark{7}, Joseph B. Gurman\altaffilmark{8}, William D. Pesnell\altaffilmark{8}, Michael J. Thompson\altaffilmark{1}}

\altaffiltext{1}{High Altitude Observatory, National Center for Atmospheric Research, P.O. Box 3000, Boulder, CO 80307, USA.}
\altaffiltext{2}{School of Earth and Space Sciences, Peking University, Beijing, 100871, China.}
\altaffiltext{3}{Department of Physics, Montana State University, Bozeman, MT 59717, USA.}
\altaffiltext{4}{School of Physics and Astronomy, University of Birmingham, Edgbaston, Birmingham, B15 2TT, UK.}
\altaffiltext{5}{Cooperative Institute for Research in Environmental Sciences, University of Colorado, Boulder, CO 80205, USA.}
\altaffiltext{6}{Lockheed-Martin Solar and Astrophysics Laboratory, 3251 Hanover St., Org. A021S, Bldg. 252, Palo Alto, CA  94304, USA.}
\altaffiltext{7}{Marshall Space Flight Center, Code ZP13, Huntsville, AL 35812, USA.}
\altaffiltext{8}{Solar Physics Laboratory, NASA Goddard Space Flight Center, Greenbelt MD 20771, USA.}

\begin{abstract}
Sunspots are a canonical marker of the Sun's internal magnetic field which flips polarity every $\sim$22-years. The principal variation of sunspots, an $\sim$11-year variation in number, modulates the amount of magnetic field that pierces the solar surface and drives significant variations in our Star's radiative, particulate and eruptive output over that period. This paper presents observations from the Solar and Heliospheric Observatory and Solar Dynamics Observatory indicating that the 11-year sunspot variation is intrinsically tied it to the spatio-temporal overlap of the activity bands belonging to the 22-year magnetic activity cycle. Using a systematic analysis of ubiquitous coronal brightpoints, and the magnetic scale on which they appear to form, we show that the landmarks of sunspot cycle 23 can be explained by considering the evolution and interaction of the overlapping activity bands of the longer scale variability.
\end{abstract}

\section{Introduction}
Early investigations of the enigmatic spots on the Sun revealed that their number waxes and wanes over a period of about 11 years \citep[e.g.,][]{1844AN.....21..233S} Ð a phenomenon that became known as the sunspot (or solar) cycle \citep[e.g.,][]{1904MNRAS..64..747M}. It was subsequently found that the latitudinal distribution of sunspots, and their progression over their evolutionary cycle, followed a trail from mid solar latitudes (about $\pm$35\degr) at first appearance, through solar maximum (when their number is at its greatest) to their eventual disappearance near the equator (about $\pm$5\degr) into the relative calm of solar minimum. Following minimum, a couple of years later, the spots appear again at mid-latitudes and the progression to the equator starts afresh, defining the start of the next sunspot cycle. 

The pattern that sunspot locations make in this cyclic progression when latitude is plotted versus time is dubbed the ``butterfly diagram'' and has become an iconic image of the Sun's variability \citep[e.g.,][]{2010LRSP....7....1H}. Continuing this rapid pace of discovery, G.~E. Hale and colleagues subsequently identified that sunspots were locations of intense magnetic field \citep[][]{1919ApJ....49..153H} and that, in consecutive butterfly wings (sunspots in the same-hemisphere but the belonging to the next cycle), the sunspots had opposite magnetic polarities \citep[][]{1924Natur.113..105H}. Indeed, they had discovered that the sign of the prevalent magnetic field in each hemisphere of the Sun undergoes a complete period every 22 years \citep[e.g.,][]{1959ApJ...130..364B,1992ASPC...27..335H}. 

The radiative and particulate output of the Sun is strongly modulated by the 11-year sunspot cycle. The continued observation and cataloging of sunspots since the pioneering observations of Schwabe, Maunder, and Hale have provided us with the most striking proxy of the Sun's activity cycle which induces both violent (short-term \-- ``Space Weather'') and gradual (long-term \-- ``Space Climate'') changes in the Sun-Earth connection. The ever-increasing reliance of humanity on space-based technology has reached the point where understanding the origins and impacts of magnetic activity of our star is imperative.

In the following sections we present an analysis of small ubiquitous features in the Sun's photosphere and corona that, when identified in images and tracked over a period of time, illustrate a considerably longer record of systematic magnetic evolution than sunspots. Further, the observed temporal progression appears to illustrate the latitudinal variation of oppositely signed toroidal magnetic flux systems, or ``bands'', belonging to Hale's 22-year magnetic polarity cycle. This observational finding confirms the observational work of \citet{1988Natur.333..748W} and \citet{1992ASPC...27..335H} who were first to identify the observational traces of an ``extended solar cycle'' \citep[see also][]{2013SoPh..282..249T}. Finally, we expand upon our analysis, and that of these pioneering papers, to show that the landmarks of sunspot cycle 23 can be phenomenologically described by considering the latitudinal interaction between these overlapping longer-lived bands.

\section{Observations and Analysis}\label{method}
A map of the magnetic range of influence \citep[or ``MRoI'';][]{2006ApJ...644L..87M} is constructed from a line-of-sight magnetogram in a pixel-by-pixel fashion and is defined as the (radial) distance from that pixel at which the total signed flux of the enclosed region is zero. The MRoI is a measure of magnetic balance, or the effective length scale over which we would expect the overlying corona to be connected, or closed. 

\begin{figure}
\epsscale{0.75}
\plotone{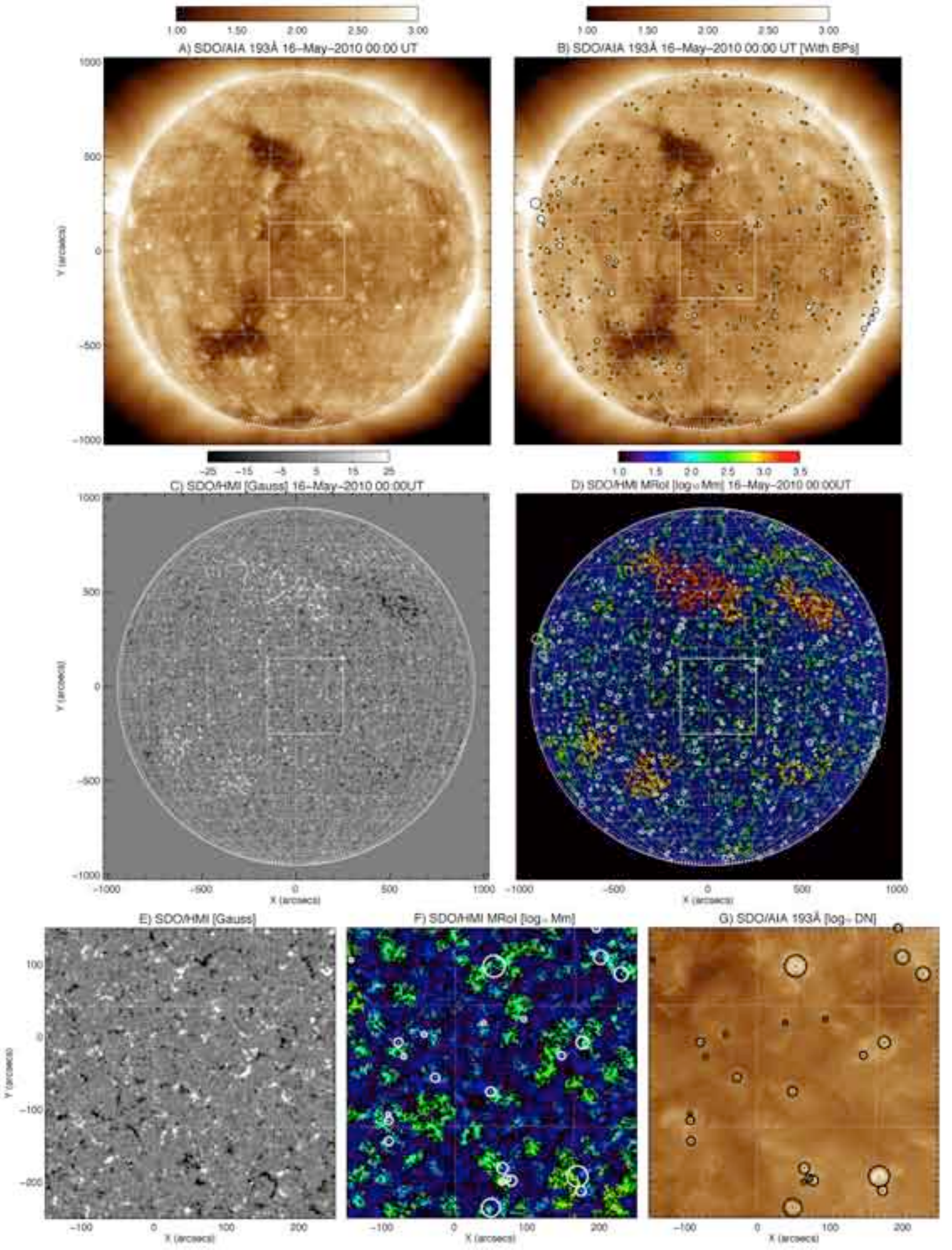}
\caption{Adapted from \citet{McIntosh2014}. EUV Brightpoints (BPs) and surface magnetism. Full disk \sdo{}/AIA image of the solar corona in the 193\AA{} channel (A). Panel B shows the locations of the detected coronal BPs (black circles). Panel C shows the corresponding full-disk \sdo{}/HMI line-of-sight magnetogram while panel D shows the derived ``Magnetic Range of Influence'' map. The MRoI map also has the coronal BP locations marked (white circles). Comparing the inset regions of panels A through D (white squares) the \sdo{}/HMI magnetogram (E), the MRoI (F), and the coronal environment in \sdo{}/AIA 193\AA{} (G).  \citet{McIntosh2014}noted the strong spatial correspondence between the BP location and the concentrations of MRoI which display a 100-250Mm  connective length-scale. \label{f1}}
\end{figure}

Figure~\pref{f1} illustrates correspondences between \sdo{}/HMI MRoI and coronal structures. \citet{McIntosh2014} discussed the four apparent length scales that are present in a typical MRoI map like this: a scale at the resolution limit of \sdo{}/HMI peaking at $\sim$5Mm which may be a signature of granules; a very large contribution peaking at $\sim$25Mm that is consistent with the mean size of quiet sun supergranules, a distribution of 100-250Mm scale objects that is consistent with the spatial dimension of giant convective cells \citep[see, e.g.,][]{2005LRSP....2....1M}; and the very long connective length scales of active regions and coronal holes for which the method was originally conceived \citep[][]{2006ApJ...644L..87M, McIntosh2007a, McIntosh2010}. The interested reader is encouraged to follow the discussion of  \citet{McIntosh2014} for further detail on the method's determination of the giant convective scale, but state that this rotationally driven global convective scale offers insight into the evolution of magnetic fields that reach possibly to the very bottom of the solar convection zone. \citet{McIntosh2014} also expanded on a correspondence that was first noted by \citet{2007ApJ...670.1401M} in that EUV Brightpoints \citep[e.g.,][]{1973SoPh...32...81V} tend to recurrently form in the vicinity of the concentrations of the 100-250Mm scale, features that they dubbed ``g-nodes'' because of their apparent connection to the giant convective scale.

The circles shown in Fig.~\pref{f1}B are the locations of the EUV BPs detected in the \sdo{} Atmospheric Imaging Assembly \citep[AIA;][]{2012SoPh..275...17L} 193\AA{} imaging channel. Details of the BP detection and tracking algorithms are available in the literature \citep[see, e.g.,][]{2005SoPh..228..285M, 2003ApJ...589.1062H} although we have taken steps to improve the reliability of the detection in subsequent years. Following image calibration and cosmic ray removal (using mission-provided software) we construct a ``background'' image ($I_{b}$) using a 40Mm$\times$40Mm smooth version of that image ($I$). BPs are defined as spatially small (2-20Mm) three-$\sigma$ enhancements of the original image over the background image. That is, we construct a ``sigma'' image ($(I - I_{b}) / \sqrt{I_{b}}$). A sigma image constructed in this way can account for subtle differences from instrument to instrument, e.g., from {\em SOHO}/EIT and \sdo/AIA and provides significantly more robust BP determination (McIntosh et~al. 2014, in preparation). From the resulting histogram of sigma image values we use one, two, and three standard deviations above the mean value to identify the thresholds for BP detection \-- the three-$\sigma$ detections being the most reliable. After defining the BP detection thresholds we isolate contiguous pixel groups in the images, computing their center position and radius of gyration. Only three\--$\sigma$ regions with radii between 2 and 20Mm are defined as belonging to BPs and in the case presented we make no effort to separate the quiet and active region BPs. The use of three\--$\sigma$ BPs (as we have in this paper) allows a large degree of confidence in the detection and latitudinal variation of the BPs. Panels A and B of Fig.~\pref{f1} show an \sdo/AIA 193\AA{} image of the solar corona without (A) and with (B) the BPs identified respectively. The same detection and identification methodology can be applied to the full-disk MRoI maps (panel D) to identify only the spatial concentration of 100-250Mm scale, the g-nodes. 

\begin{figure}
\epsscale{1.00}
\plotone{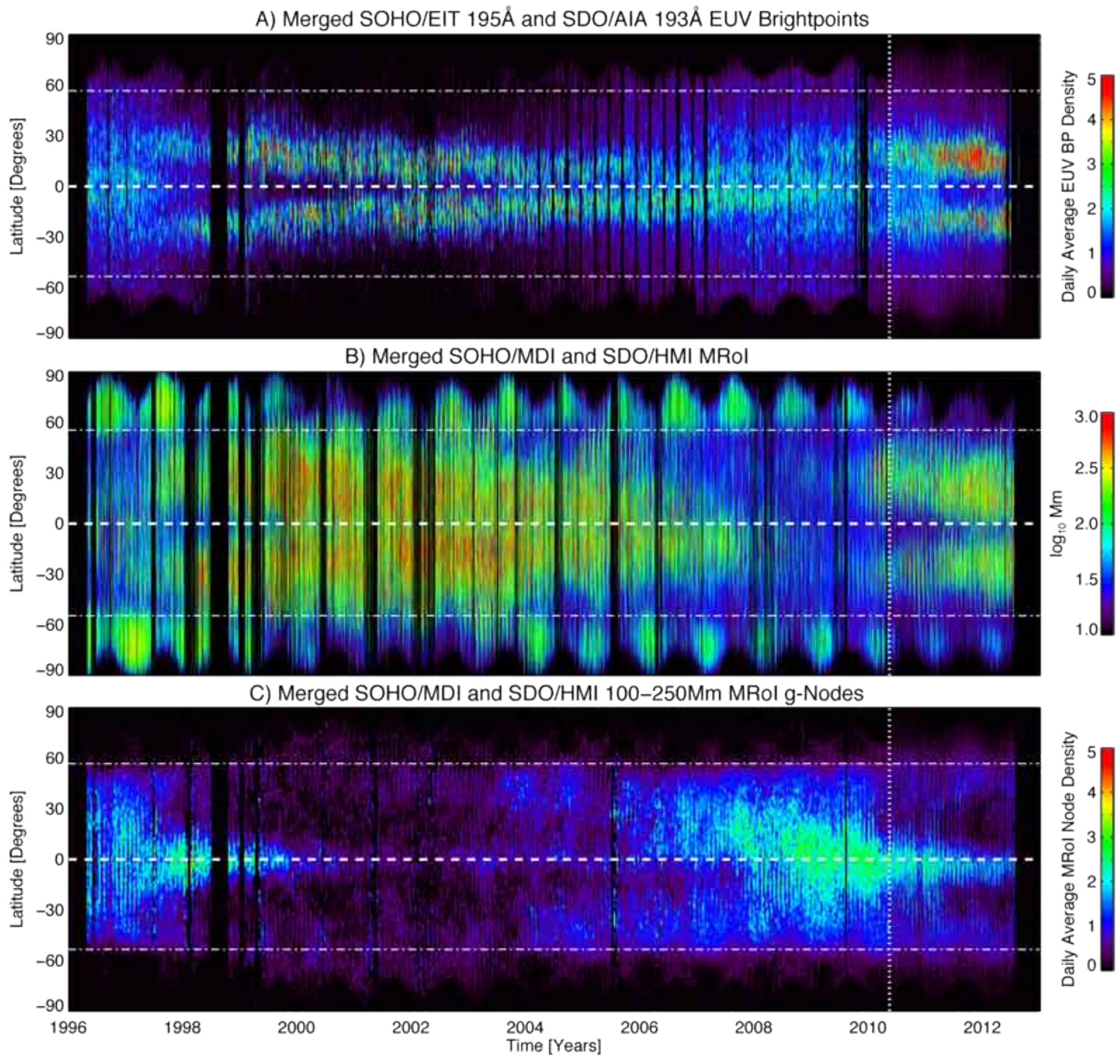}
\caption{(A) Latitude versus time plots of the merged \soho/EIT, \sdo/AIA BP locations, (B) the \soho/MDI and \sdo/HMI MRoI, and (C) the locations of only the g-nodes. The white dashed and dot-dashed horizontal lines are drawn at the equator and $\pm$55\degree{} latitude for reference. The white vertical dotted line in May 2010 indicates the transition from \soho{} to \sdo{} diagnostics. \label{f2}}
\end{figure}

We repeat the BP and g-node identification process for the \soho/EIT \citep[][]{1995SoPh..162..291D}, \sdo/AIA, \soho/MDI \citep[][]{1995SoPh..162..129S} and \sdo/HMI \citep[][]{2012SoPh..275..207S} image and magnetogram archives and the results are shown in Fig.~\pref{f2} as latitude-time plots, sampling only the variation $\pm$5\degree from the central meridian. These measures are compensated for the variation in the inclination of the Sun's poles relative to the Sun-Earth line (note the sinusoidal nature of the highest latitude regions). Panel A shows the latitude\--time variation of the EUV BP density in the \soho/EIT 195\AA{} and \sdo/AIA BP 193\AA{} images \citep[the interested reader is pointed to the figures of][for reference]{2005SoPh..228..285M}. We note a paucity of BPs detected in this fashion above $\pm$55\degree{} latitude and the appearance of multiple ``bands'' of BPs in the 1996-1998 and 2006\--2011 timeframes. The latitudinal variation of the (central meridian) MRoI is shown in panel B where perhaps the most striking pattern is the clear delineation of evolution above and below 55\degree{} latitude. Below 55\degree{} latitude the MRoI pattern evolves towards the equator \-- like the BPs \-- but the region above does not. Above 55\degree{} latitude there is little migration and the pattern appears to evolve in place. There are two excursions in the MRoI starting in $\sim$1999 in both hemispheres but the injection of short length scales in the northern hemisphere appears to end earlier ($\sim$2002) than in the southern hemisphere ($\sim$2003). Panel C shows the complex latitude\--time evolution of the g-node density. We note that the dark patches in the g-node density distribution are an artifact of the MRoI technique in that it is often difficult to see scales of order 100-250Mm inside active region and coronal hole which have spatially distributed regions of large MRoI (see panel B and Fig.~\pref{f1}D). Like the BPs there is a significant reduction in the number of g-nodes above 55\degree{} latitude and there also appear to be low density, asymmetric bands of g-nodes that reach to 55\degree{} at apparently different times, approximately 2002 in the north and 2004 in the south. Furthermore, at the same time as the multiple overlapping bands of BPs (1996-1998 and 2006\--2011) there are large concentration of g-nodes in the region below 55\degree{}.

\begin{figure}[h!]
\plotone{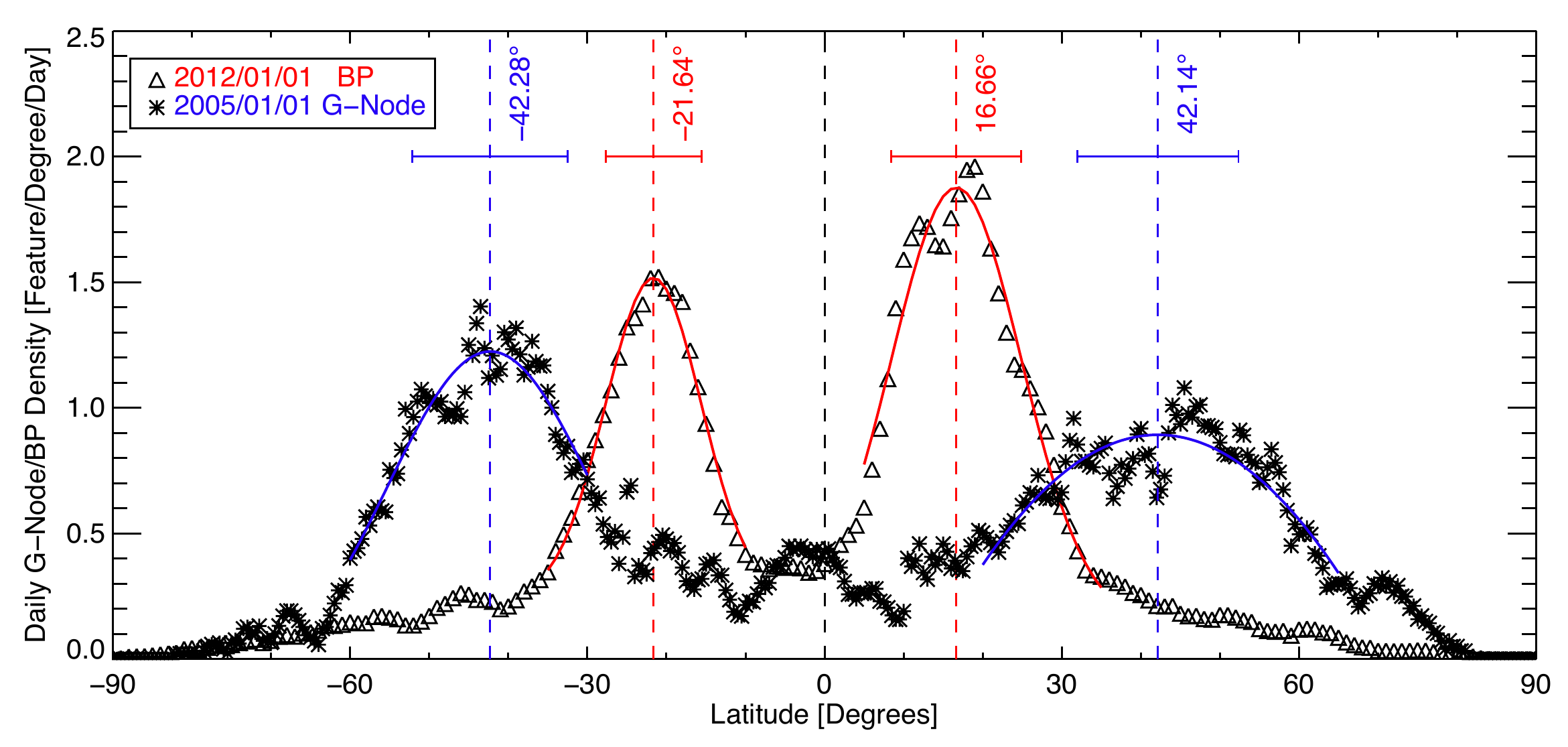}
\caption{BP (triangle) and g-node (asterisk) distributions taken from Fig.~\pref{f2} centered at different times during the time period studied. The peaks in the distributions are fitted using Gaussians which are shown as thick red and blue solid lines for BP and g-nodes respectively. The colored vertical dashed lines indicate the latitude of the relevant distributions peaks while the horizontal bars show the Gaussian sigma for the distribution fit. \label{f3}}
\end{figure}

\subsection{Identifying \& Fitting Activity Bands}\label{sec-fit}
\begin{figure}[h!]
\plotone{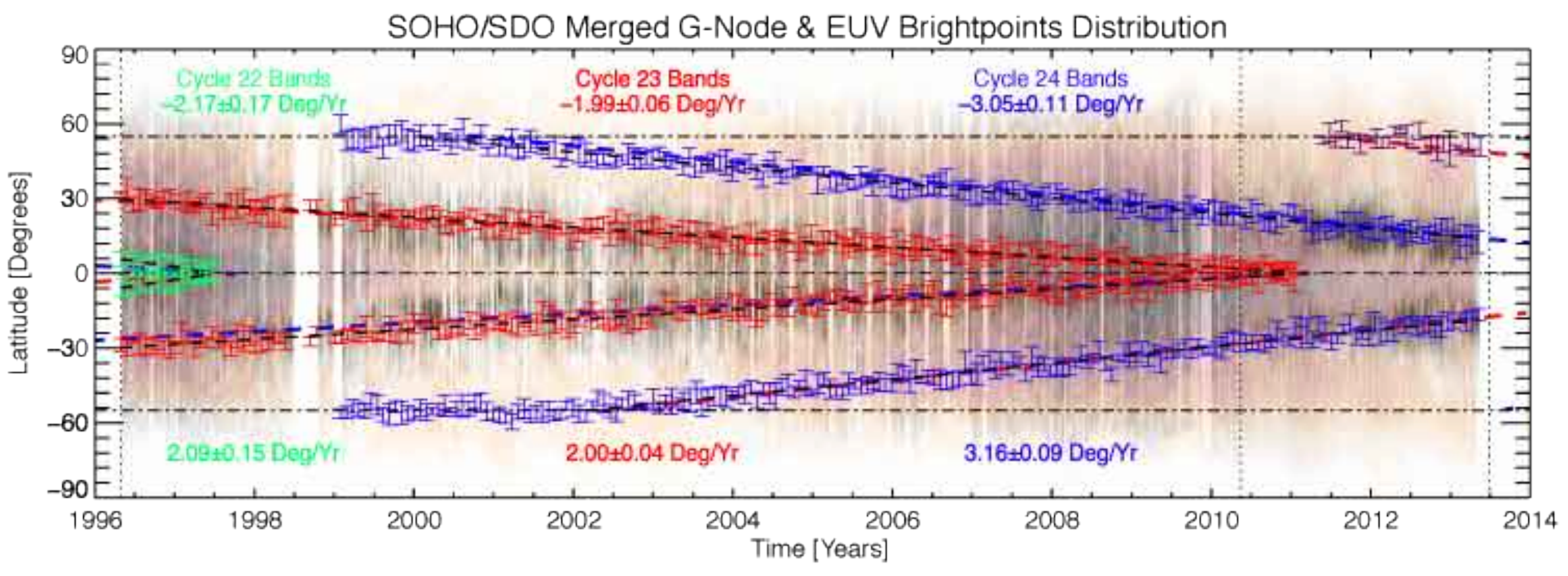}
\caption{Fitting the BP and g-node bands terminating in 1997 (green), 2011 (red), and the current bands (blue) from the combined g-node and BP latitude-time distributions. Each bar on the plot is determined as shown in Fig.~\pref{f3} and the results are assumed to describe a linear migration of the activity band with time. The linear fit to each band is shown as a black dashed line and the gradients fit are as shown on the plot. The vertical dotted lines mark the beginning and end of the observation sample. \label{f4}}
\end{figure}

To quantify the discussion of Fig.~\pref{f2} we use the daily histograms of BP and g-node density to identify and track the latitudinal migration of these features. We use the following recipe to map the positions of the activity bands in the latitude-time figures:
\begin{enumerate}
\item{Averaging the latitudinal BP and g-node distributions over a running 28\--day window we construct histograms of each as a function of latitude like those shown in Fig.~\pref{f3} where we draw comparison with those published previously \citep[e.g.,][]{1978ApJ...219L..55G}.}
\item{We find the latitudinal location of the histogram peaks, allowing for a maximum of four (two per hemisphere). The histogram peaks are fitted using Gaussians from which we can determine their mean latitude and width with the latter being employed as an (upper) estimate of the error in establishing the mean latitude. The fits are manually verified, the peak locations are sorted by latitude and associated with a particular activity band for the next step. Figure~\pref{f3} shows example fits for g-node and BP distributions at two distinct time steps noting that in the seven years separating the samples these bands have traveled from approximately 42\degree{} to $\sim$20\degree{} in each hemisphere. As the bands migrate closer to the equator it becomes more difficult to isolate them and will associate the same peak and width with one band in each hemisphere.}
\item{After repeating step 2 for all times we use the peak locations and widths associated with each band to map their latitudinal progression. A straight line is chosen to fit these points for each band under the assumption that it is the functional form that is minimally consistent with the data points. The results of this analysis are shown in Fig.~\pref{f4}.}
\end{enumerate}

From the analysis of Fig.~\pref{f4} it appears that the migrating bands of g-node and BPs which we will refer to hereafter as ``activity bands,'' appear to meet, or terminate, at the equator. The three different sets of bands can be associated with solar cycles 22 (green), 23 (red) and 24 (blue). For the currently visible portion of the solar cycle 24 activity bands the fitted migration gradients are consistent with each other (3.05±0.11\degree/yr in the northern hemisphere and 3.16$\pm$0.09\degree/yr in the south) within the uncertainty in the fit. These values are higher than, about 1 degree per year higher, than the values for the portions of the cycle 22 (green bars) and cycle 23 (red bars) activity bands shown and indicates that the linear approximation describing the activity band motion is not the most accurate and that the bands may slow down as they approach the equator.  Note that the purple points, occurring at high latitudes in the northern hemisphere in late 2011, they indicate the possible start of the cycle 25 activity band. If these bands do connect to the sunspot formation bands they last significantly longer that the $\sim$11 years on which sunspots are visible in each cycle (see below) and would appear to slow down as the cycles come to an end. 

\begin{figure}[h!]
\epsscale{1.00}
\plotone{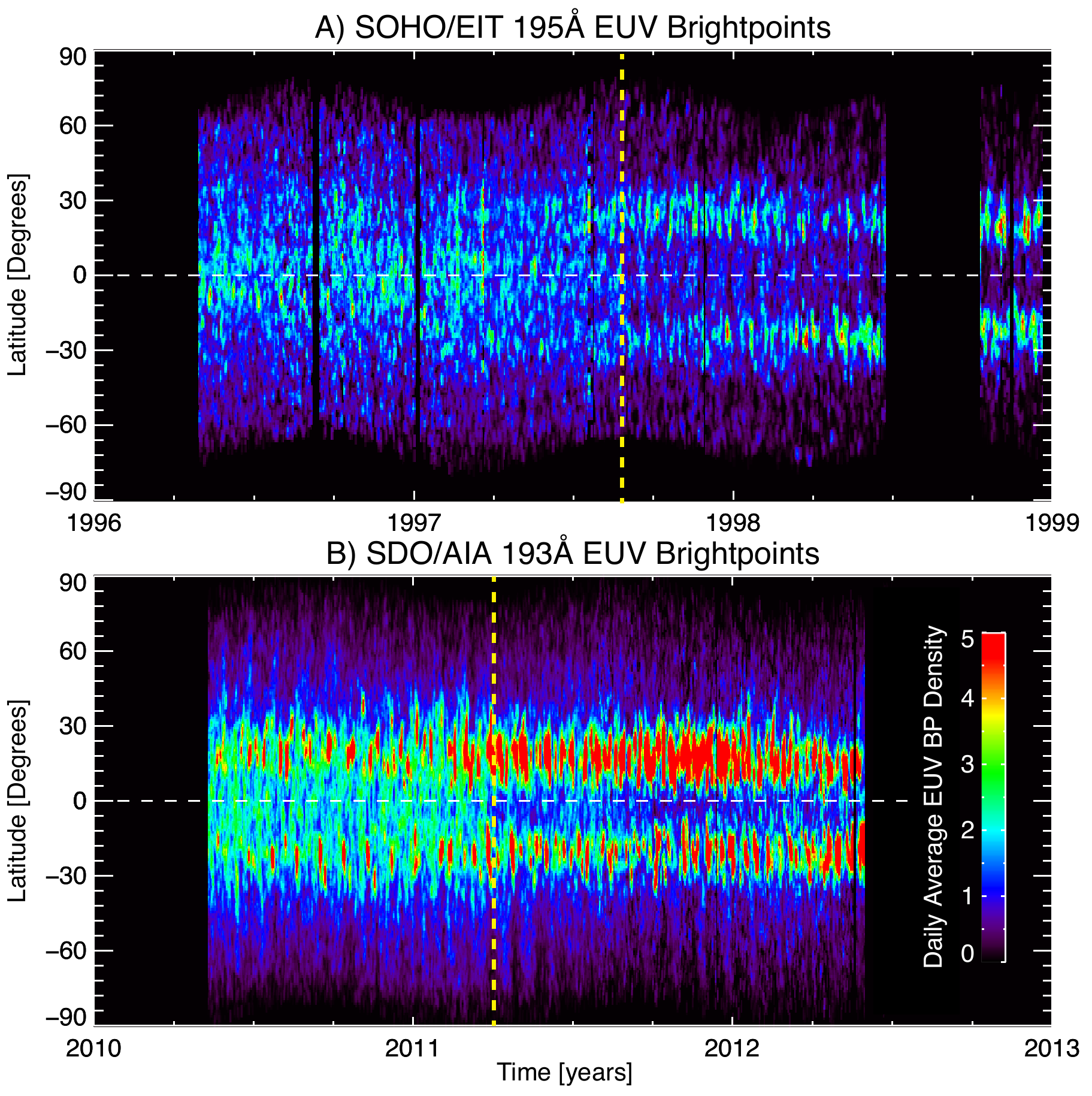}
\caption{Latitude-time BP distributions for \soho/EIT 195\AA{} through the solar cycle 22/23 minimum (1997-1999; A) and the solar cycle 23/24 minimum (2010-2012; B) for \sdo/AIA 193\AA{}. In these panels we see bands coming to an end in August 1997 and March 2011 respectively. The latitudinal variation in BPs shows a marked change following termination at the times marked with vertical dashed yellow lines. \label{f5}}
\end{figure}

To illustrate, and validate, the apparent termination points Fig.~\pref{f5} shows the \soho/EIT and \sdo/AIA BP distributions in the 1996 \-- 1999 (panel A) and 2010 \-- 2013 (panel B). In each case we see that the latitudinal distribution of BPs following is starkly different after August 1997 and March 2011 respectively.

\subsection{The Space-Time Progression of Small-Scale Features}
\begin{figure}
\epsscale{0.9}
\plotone{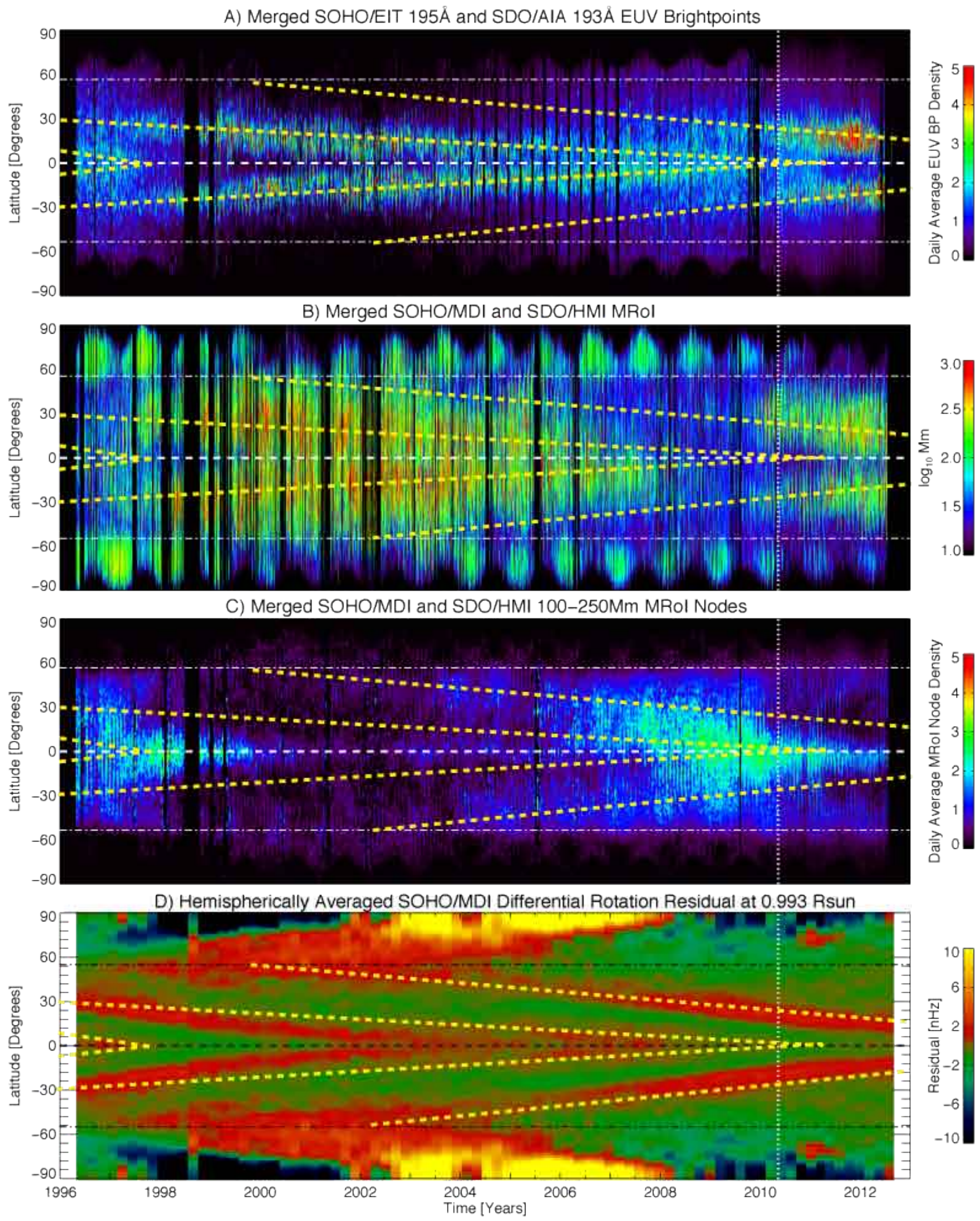}
\caption{(A) Latitude-time plots of the merged \soho/EIT, \sdo/AIA BP density distribution, (B) the \soho/MDI and \sdo/HMI MRoI, (C) the g-nodes density distribution, and (D) the (hemispherically-symmetrized) torsional oscillation near the solar surface \citep[][]{2000Sci...287.2456H}. The dot-dashed horizontal lines are drawn at $\pm$55\degree latitude and separate the polar and equatorial regions. The yellow dashed bands are drawn using a combination of panels A, and C (as explained in the text and illustrated in Fig.~\pref{f4}) using only the pieces that migrate in latitude. \label{f6}}
\end{figure}

Figure~\pref{f6} extends the analysis of BPs, g-nodes and the MRoI over the seventeen year record of synoptic coronal and photospheric observations accrued by the \soho/EIT, \soho/MDI, \sdo/AIA and \sdo/HMI (cf. Fig.~\pref{f2}). Panels A, B and C repeat those shown in Fig.~\pref{f2} however, now we place the activity bands inferred from the g-node and BP density histograms (Fig.~\pref{f4}) on top of the latitude-time plots. For further comparison we show the hemispherically averaged residual to solar differential rotation inferred from MDI and HMI Dopplergrams at a depth of 0.993Rsun in panel D \citep[][]{2009LRSP....6....1H}. This reveals a pattern known as the ``torsional oscillation'' \citep[e.g.,][]{2010LRSP....7....1H, 1982SoPh...75..161L, 1999ApJ...526..523C} or a measure that has been interpreted as a map of the large-scale zonal flows in the solar interior \citep[e.g.,][]{1996ApJ...460.1027H, 2010ApJ...725..658U}. 

Again, now assisted by the linear fits to the activity bands developed in Sect.~\pref{sec-fit}, the latitudinal progression towards the equator with time is clearly visible in all quantities plotted in the panels of the figure. We immediately notice the (general) correspondence between the surface magnetism inferred activity bands and the torsional oscillation but, due to the hemispheric averaging of the latter, some departures are visible. The equatorward migration of BPs, g-nodes and their link to the torsional oscillation provides further evidence of deep rooting of these magnetic field concentrations \citep[][]{McIntosh2014}, otherwise one would expect their motion to be poleward with the flux elements caught in the poleward meridional circulation \citep[e.g.,][]{2005LRSP....2....5S,2010LRSP....7....3C}. We see again that that the dot-dashed lines at $\pm$55\degree{} latitude divide the poleward and equatorward behavior, something that is especially clear in panel D  where it is the latitude at which the torsional oscillation pattern diverges with one pattern going poleward and another going equatorward. The importance and relevance of $\pm$55\degree{} will be discussed in a later section.

We see that there is strong hemispheric asymmetry in the magnetic activity of solar cycle 23 \citep[as is the focus of][]{2013ApJ...765..146M}. In the present case we see that the small-scale magnetic activity bands start their equatorward migration from high latitudes asymmetrically. After appearing in 1999 (seen as horizontally extended clusters of g-nodes and in the torsional oscillation at high latitude) the northern band started migrating from 55\degree{} in 2000 while the southern band started migrating later \-- some time in 2002. It is unclear why the bands, although they appear at the same time, would start migrating equatorward with such a significant offset but the responsible physical mechanism is likely the root cause of the the current hemispheric asymmetry in activity \citep[][]{2013ApJ...765..146M}. With the aid of the activity band linear fits (Fig.~\pref{f4}) we can establish that there are relatively short periods of time when there is only one migrating activity band visible in each hemisphere.Much of the time the system in each hemisphere appears to have overlapping bands with, as we have noted earlier, the overlap most easily visible in the BP latitudinal distributions between 1996 and 1998, and 2006 and 2011, i.e., during the solar cycle 22/23 and 23/24 solar minima \citep[e.g.,][]{2013ApJ...765..146M}. Hereafter, we will refer to the combination of the hemispheric activity bands across the solar equator as forming an ``chevron''.

The combined data sample presented herein spans only seventeen years, the activity bands visible in the latitude-time plots must exist considerably longer than the $\sim$11-year sunspot cycle and we deduce that they are more closely related to the $\sim$22-year magnetic activity cycle. For example, the high latitude chevron (starting in 2000) has progressed to a latitude 20\degree{} in approximately 13 years, a latitude reached by the lower latitude chevron in 2002 approximately 9 years before it disappeared. The present analysis of BP and g-node evolution solar cycle 23 (and the start of cycle 24) is strengthened by considering the pioneering observational work of \citet{1988Natur.333..748W} and \citet{1992ASPC...27..335H} who studied BPs,``ephemeral active regions'', and bright Calcium faculae over solar cycles 19 through 21 with \citet{1988Natur.333..748W} including the torsional oscillation for cycle 21. {\em Our analysis both reproduces and confirms the earlier analyses to the degree that, when combined, we have directly observed a systematic progression of small-scale magnetic flux from high to low latitudes that spans close to five sunspot cycles.}

\begin{figure}
\epsscale{0.75}
\plotone{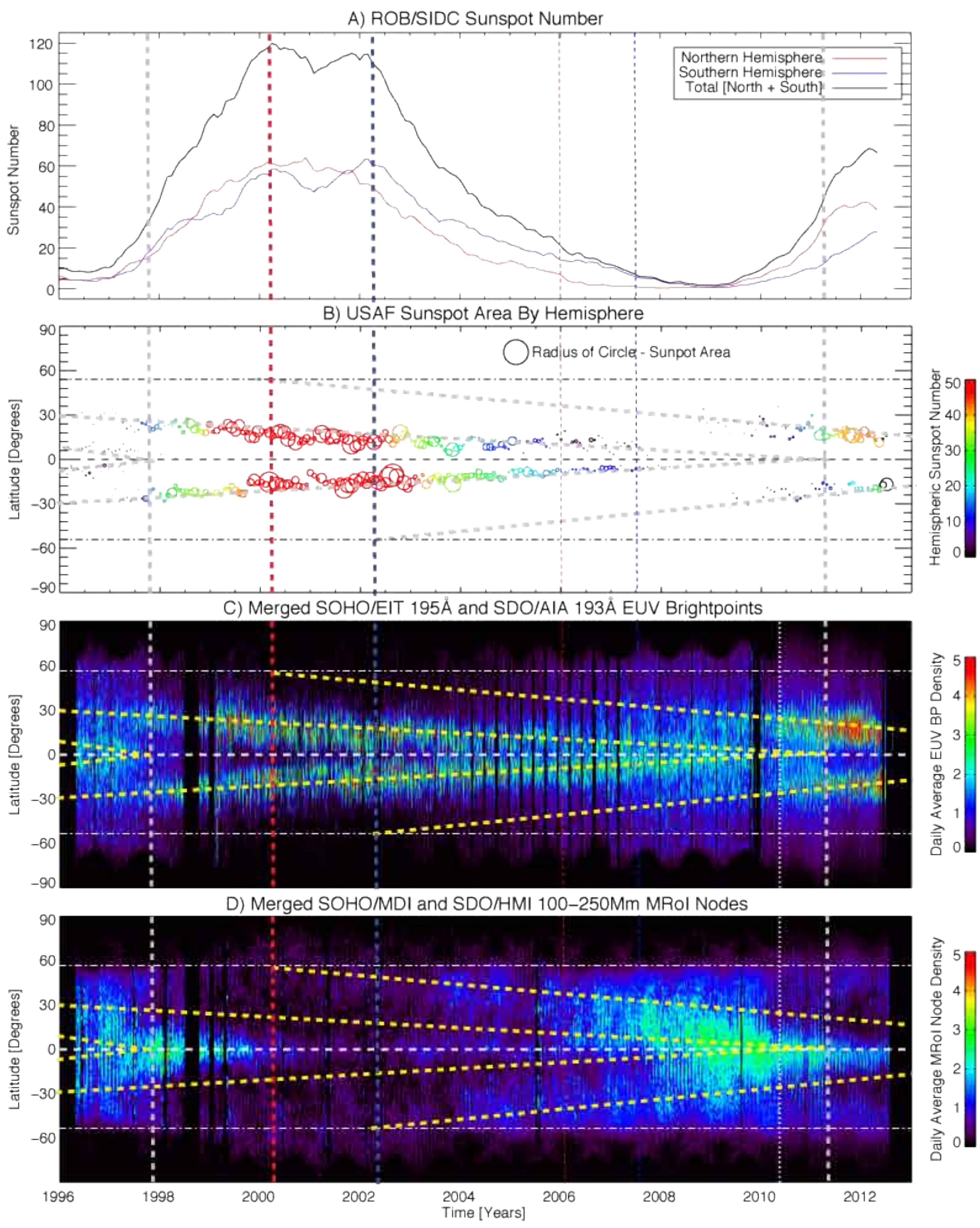}
\caption{Highlighting the phases of solar cycle 23 and the initial phase of solar cycle 24. Panel A shows the Royal Observatory of Brussels (ROB) Solar Influences Data Center (SIDC) total SSN and the SSN decomposed by hemisphere (hSSN; north Ð red; south Ð blue). Panel B shows the latitudinal variation of the SSN and sunspot areas from the United States Air Force Sunspot area archive (B). The circles drawn are color-coded by the hSSN and their radii reflect the area of the solar disk (in millionths) covered by sunspots averaged over a 28-day period. Panels C and D are reproductions of the BP and g-node latitude-time plots in Fig.~\pref{f6} and the chevrons in dashed yellow lines are also taken from Fig.~\pref{f6}. The red, blue, and grey dashed vertical lines mark the hemispheric sunspot maxima in the north, south and the apparent activity band termination points (see text and fig.~\pref{f5}. The white vertical dotted lines indicate the transition from SOHO to SDO diagnostics in May 2010. \label{f7}}
\end{figure}

\section{Comparing Activity Band Progression With The Hemispheric Sunspot  Number}
To go beyond the analysis of \citet{1988Natur.333..748W} \citep[and][]{1992ASPC...27..335H} we now explore the relationship between the activity bands and the modulation of the sunspot cycle by placing them in context with the variation in the (monthly) hemispheric sunspot number (hSSN) and the \url[http://solarscience.msfc.nasa.gov/greenwch.shtml]{United States Air Force (USAF) sunspot archive} in Fig.~\pref{f7}. The thick grey, red, and blue vertical dashed lines in the figure are landmarks of sunspot cycle 23. The grey lines mark the times at which the low latitude chevrons appear to terminate at the equator (in August 1997 and April 2011; Fig.~\pref{f5}). 

For the two terminations visible we see that the sunspots of the upcoming cycle appear rapidly, with great abundance and increasing strength on the remaining activity band in each hemisphere after the termination. Incidentally this transition happens as the higher latitude band passes $\sim$30\degree{} latitude. We infer from these examples that the grey lines define the start of the ascending phase of sunspot cycles 23 and 24. The thick dashed red and blue vertical lines mark the asymmetric times when the high ($\pm$55\degree) latitude bands start their progression to the equator in the northern and southern hemispheres. We see that the start of these lines coincide in time with the hemispheric sunspot maxima and, hence, mark the start of the declining phase in each hemisphere. 

The thin red (January 2006) and blue (September 2007) dotted lines mark when the higher latitude activity bands passes 45\degree{} \-- this starts a period of time when all four activity bands begin to overlap at low latitudes. At these times the sunspots in each hemisphere rapidly begin to wane in size and number. This period of maximum overlap coincides with the activity/sunspot minimum between solar cycles 23 and 24, a state which appears to persist until the next grey dashed line when the low-latitude bands terminate.

\begin{figure}[!h]
\epsscale{1.00}
\plotone{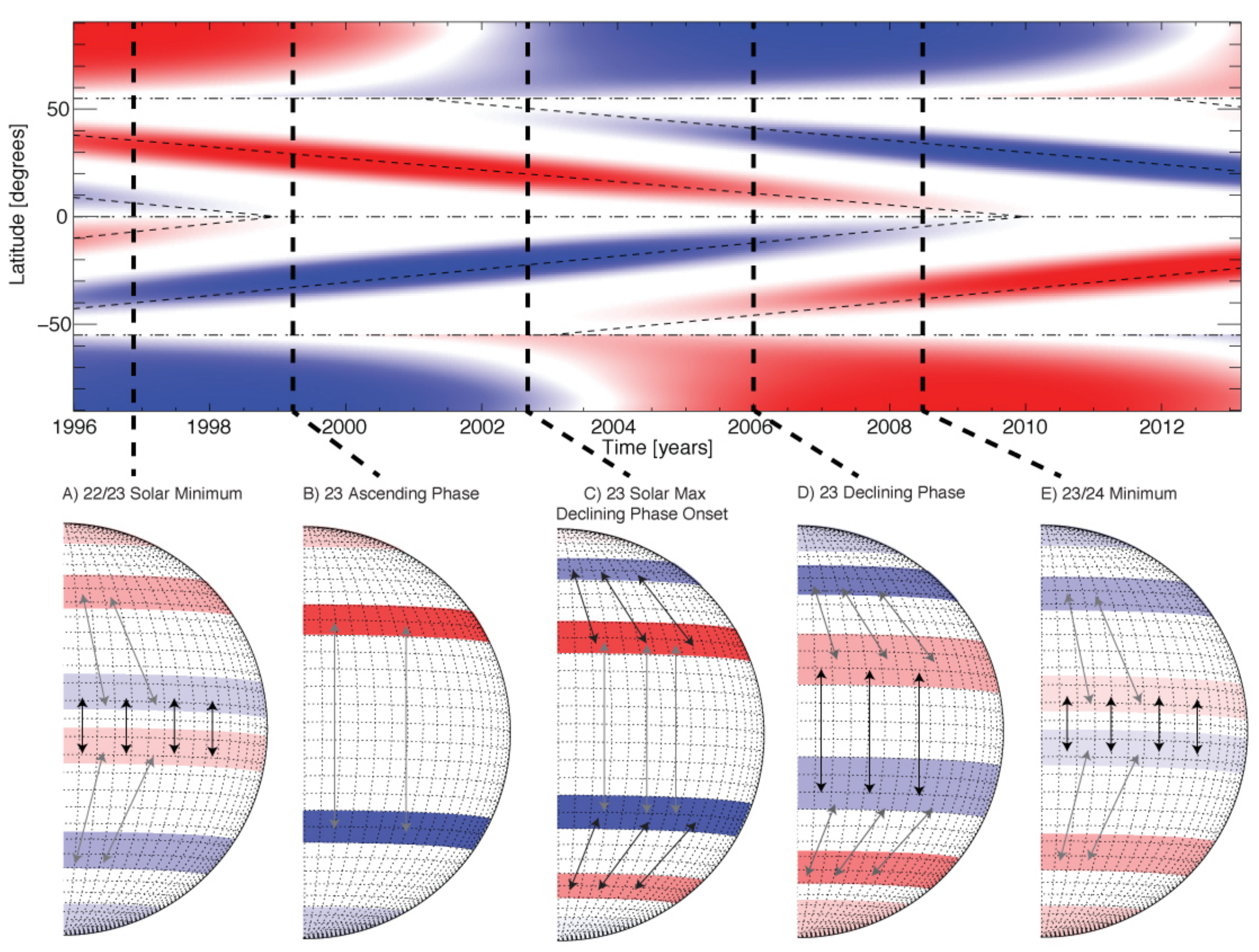}
\caption{The progression of the sub-surface toroidal magnetic activity bands and polar caps as inferred from the analysis of Figs.~\pref{f6} and~\pref{f10}. From left to right: the minimum between solar cycles 22 and 23 (A, $\sim$1996), the ascending phase (B, $\sim$1999), through polar reversal and solar maximum (C, $\sim$2002), into the declining phase (D, $\sim$2004), and finally to the minimum between solar cycles 23 and 24 (E, $\sim$2009). The red and blue bands indicate the presence of positive and negative polarity magnetic fields respectively where deeper color indicates stronger field. The arrows are drawn between the bands to indicate the magnetic connections between the bands in their hemisphere and across the equator.} \label{f15}
\end{figure}

Figure~\pref{f15} illustrates our interpretation of the cycle 23 activity bands in terms of the underlying magnetism and its variation with time. At the start of the \soho{} era (the activity minimum of 1996/1997) the net signed flux of the four bands of opposite sign cancel each other within their own hemisphere and across the equator. The termination of the low-latitude bands in 1997 leaves only the single activity band in each hemisphere. The cancelation of the oppositely signed bands across the equator significantly increases the flux density of the remaining high-latitude band in each hemisphere which then permits the rapid formation and buoyant rise of sunspots on those bands. As we approach solar maximum a new activity band starts to appear at $\sim$55\degree{} latitude. While sitting at that latitude some of the magnetic flux in that band emerges and is caught in the surface meridional flow and is advected poleward and eventually cancels the existing polar field above 55\degree. As the new activity band starts to migrate equatorward the magnetic flux that it contains begins to interact with that of the lower latitude (oppositely signed) band in the same hemisphere. The increased interaction reduces the local flux density in the low latitude bands, reducing their ability to buoyantly produce sunspots. {\em The result will be a net reduction in available flux and hence activity}. In this picture sunspot maximum in each hemisphere is the time when this intra-hemispheric interaction starts afresh. As time progresses and the four (alternating-polarity) activity bands migrate towards the equator, the oppositely signed activity bands interact across the equator as well as in the same hemisphere and the situation becomes the equatorial reflection of the previous solar minimum.

Based on the correspondence of these observational datasets we deduce the following as the landmarks of the solar (sunspot) cycle:
\begin{itemize}
\item{``Solar Minimum'' is the period of time during which the oppositely signed toroidal flux systems in each hemisphere, and across the equator, mutually cancel.}
\item{The ``ascending phase'' of the sunspot cycle involves only one activity band per hemisphere and commences with the end of solar minimum - that is, when the trans-equatorial flux systems cancel.}
\item{``Solar Maximum'' is the time at which the new high latitude flux system begins migrating towards the equator.}
\item{The ``descending phase'' of the sunspot cycle involves two activity bands per hemisphere. There is likely a complex state of sub-surface interaction between these bands.}
\end{itemize}

\section{Supporting EUV Observations of Cycle 23 Evolution}

\begin{figure}[!h]
\epsscale{1.00}
\plotone{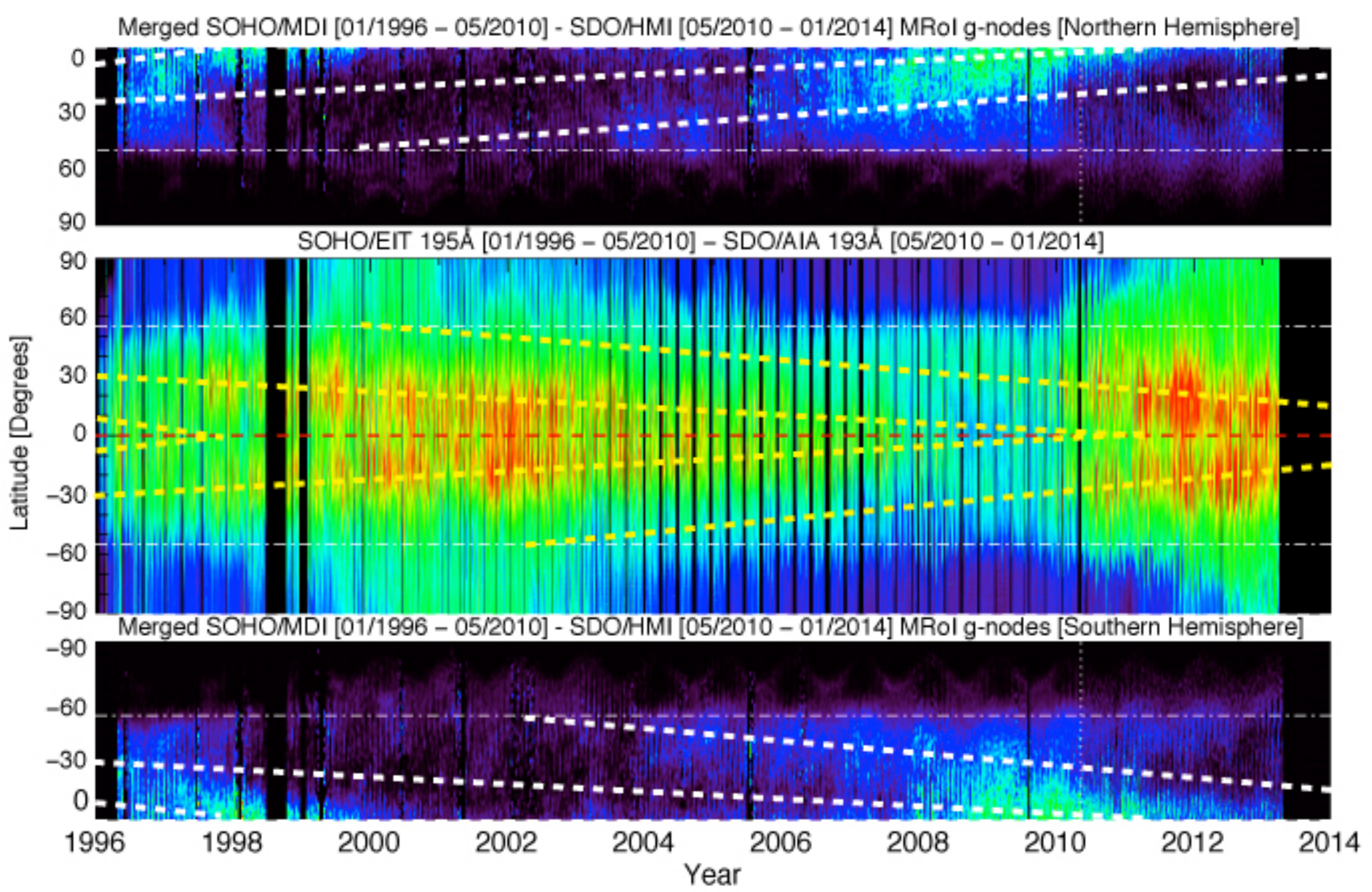}
\caption{Comparing the latitudinal progression of coronal emission around the solar limb from \soho/EIT (195\AA) and \sdo/AIA (193\AA{}) with the latitudinal progression of the g-nodes in each hemisphere (north - top; south - bottom). The central panel is created by integrating the emission in an annulus of extent 1.15\--1.25 R$_{\sun}$. The inclined dashed lines that illustrate the migratory path of the activity belts are taken from Fig.~\pref{f6}. The horizontal dot\--dashed lines mark $\pm$55\degree{} latitude. Note the correspondence in each hemisphere between the duration of the poleward surge and the length of time that the g-node activity band spends at $\pm$55\degree{} and their clear difference in duration.} \label{f8}
\end{figure}

\begin{figure}
\epsscale{0.65}
\plotone{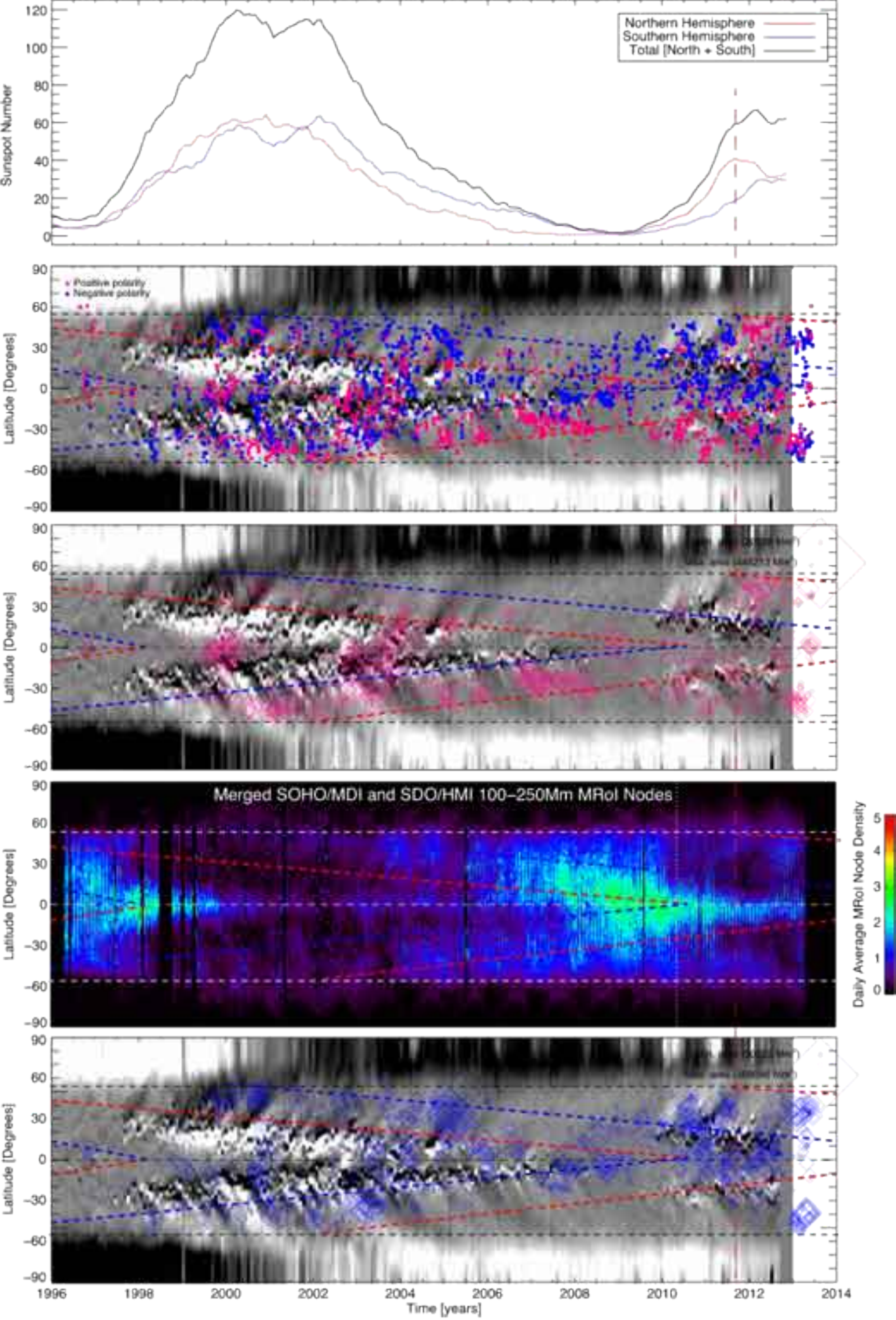}
\caption{The results of automated coronal hole detection by the CHARM algorithm using images and magnetograms from \soho{} and \sdo{} \citep[][and Krista et al. 2013 in preparation]{2009SoPh..256...87K}. From top to bottom the panels of the figure show the hemispheric sunspot number (north \-- red; south \-- blue, from Fig.~\pref{f6}), the centers of the coronal holes identified which have an area greater than 20,000Mm$^2$ (positive \-- red; negative \-- blue) overlaid on the MDI/HMI latitude-time plot, the area (indicated by size of the diamond) and location of only the positive polarity coronal holes greater than 20,000Mm$^2$ in area, the g-node density variation in latitude and time (cf. Fig.~\pref{f6}), and the area (indicated by size of the diamond) and location of the negative and positive polarity coronal holes greater than 20,000Mm$^2$ in area. In the lower four panels the horizontal dashed black lines are drawn at $\pm$55\degree, and the activity bands are colored red and blue to indicate their net polarity. The vertical dashed red line running through the entire plot indicates the time of northern hemisphere solar maximum, at time of writing the southern hemisphere had not yet reached maximum and so there is no equivalent line (see text).} \label{f9}
\end{figure}

We find further support for the BP and g-node observations of cycle 23 presented above in different analyses of coronal EUV images from \soho{} and \sdo{}: the essential nature of $\pm$55\degree, the length of time spent by the activity bands at $\pm$55\degree{}, the short-term temporal variability intrinsic to the system, and the equatorward migration of the activity bands. Figures~\pref{f8} through~\pref{f9} use different analyses techniques to increase the physical richness of the physical picture and the intrinsic coupling of small and large scale phenomena.

Figure~\pref{f8} shows the evolution of global coronal structure above the limb in the \soho/\sdo{} 195/193\AA{} images \citep[see, e.g., Fig.~7 of][]{2013ApJ...765..146M}. These images represent the temporal progression of the EUV emission integrated in a narrow (1.15\--1.25 R$_{\sun}$) annulus around the solar limb. In addition to showing the latitudinal progression of the activity bands towards the equator (the dashed lines again trace out the progression of the activity bands and are taken from Figs.~\pref{f6}) we see that the extent of the polar coronal hole very rarely migrates closer to the equator than $\pm$55\degree{}. Indeed, the polar coronal holes seem to be strongly bounded by $\pm$55\degree{}. The main poleward excursions observed are the notable poleward surges of coronal emission and polar crown filaments \-- the so-called ``Rush to the Poles'' \citep[e.g.,][]{1988Natur.333..748W,2008ASPC..383..335A,2013SoPh..282..249T} \-- near the sunspot maxima of cycle 23 (and 24 currently ongoing). 

We note that the duration of the surges appear to be identical to the length of time spent by the g-node activity bands at $\pm$55\degree{} {\em before} their migration toward the equator. The g-node progression in each hemisphere and dashed activity bands are shown above and below the central coronal plots to emphasize this correspondence. We see that the surges in the northern and southern hemisphere appear to start at the same time (in 1998) and end when the g-node bands begin their equatorward march which correspond to the clear offset between migration start in the north (2000) and south (2002) as we have noted previously. The apparent correspondence between the duration of the polar surges and the length of time spent at $\pm$55\degree{} by the (g-node) activity bands would appear to support our earlier assertion that the magnetic flux emerging from the activity band at high latitude ``feeds'' the polar region with oppositely signed magnetic flux. This flux eventually closes the polar coronal hole and reverses the net flux of the polar regions after that flux is caught in the surface (poleward) meridional flow. Indeed, as we have noted, this pattern is repeating currently indicating that the activity band representing the magnetic activity bands responsible for solar cycle 25 are present at high latitude with surges starting in 2010 and 2012 in the northern and southern hemispheres respectively.

Figure~\pref{f9} shows the evolution of coronal holes automatically detected using EUV images and magnetograms from \soho{} and \sdo{} \citep[][]{2009SoPh..256...87K} and places them in context with the evolution of the activity bands, the hemispheric sunspot number and the underlying photospheric magnetism. The panels of the figure shows the hemispheric sunspot number (north \-- red; south \-- blue, from Fig.~\pref{f6}), and coronal hole characteristics for comparison with the MDI/HMI latitude-time plot, the g-node density variation in latitude and time. In each of the lower panels we show the horizontal dashed lines at $\pm$55\degree and the progression of the activity bands with time (positive \-- red; negative \-- blue) to illustrate the strong correspondence between the progression of the activity bands and the identified coronal holes\footnote{A broader discussion of the coronal hole detection algorithm and the detailed study of this progression is presented in a subsequent paper (Krista et al. 2014 in preparation).}. 

Plotting the central location and area of the coronal holes with area greater than 20,000Mm$^2$ (chosen to represent the area covered by g-nodes with a separation of 80Mm) formed below 70\degree{} latitude and color\-coded by their underlying polarity (positive \-- red; negative \-- blue). We see that there is a general migration of the low-latitude coronal holes towards the equator on the band of the same polarity and links their evolution to the progression of BPs and g-nodes. {\em This progression indicates that a significant population of coronal holes have deep magnetic roots and are not necessarily the result of active region diffusion by surface flows} (a subject covered in Krista et al. 2014). Similarly, we can infer from this complex ``double-helix'' pattern of coronal hole locations that coronal holes will sit at $\pm$55\degree{} for prolonged periods of time as a tracer of the underlying toroidal flux band prior to equatorward migration. Such coronal holes were readily visible in the southern hemisphere from 2000 to 2002. Indeed, such coronal holes are also visible in contemporary coronal imaging (and Fig.~\pref{f4}) \-- possibly marking the start of the cycle 25 positive activity band for at +55\degree{} (see also Sect.~\pref{forecast}) which should start its progression equatorward at the northern activity maximum shown by the red dashed vertical line. We note that band in the southern hemisphere (negative at -55\degree{}) may be starting to show, but we will have to wait and see if the southern activity maximum has been reached.

\section{Applying Cycle 23 ``Rules'' to a Longer Timeframe}
\begin{figure}[!h]
\epsscale{1.00}
\plotone{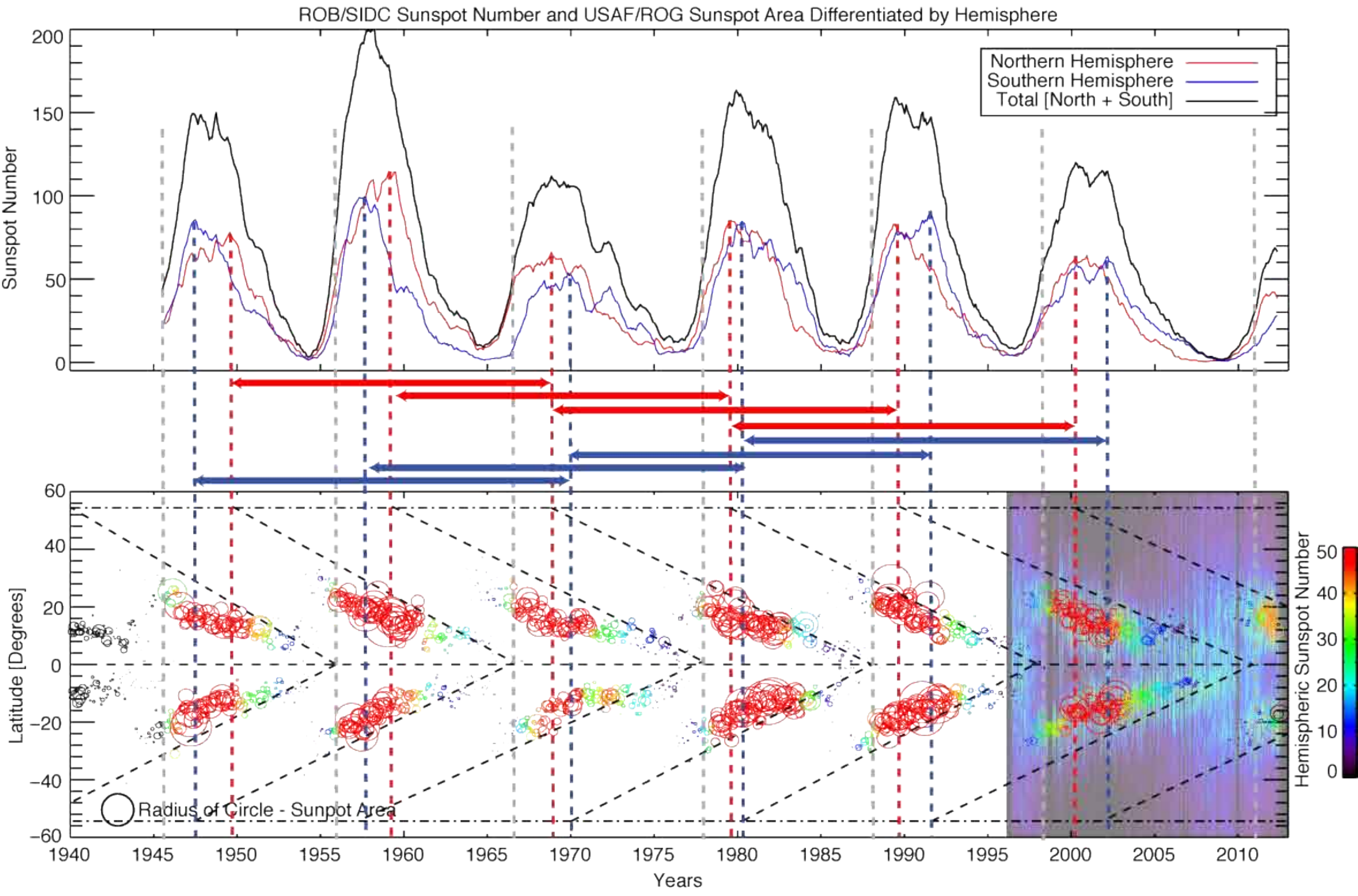}
\caption{Extending the cycle 23 metrics to a longer epoch covering 6 complete sunspot cycles (1940 \-- 2012). We use a combination of the hSSN and the USAF/Royal Observatory Greenwich record of sunspot areas. The dashed lines are drawn from the hSSN maxima (red \-- north; blue \-- south) and signify the beginning of the high latitude bands formation and migration towards the equator from $\pm$55\degree. The gray dashed lines are drawn when the sunspot areas exceeds 100 millionths, indicating that the equatorial bands have cancelled, permitting rapid high-latitude sunspot onset. The span from the colored dashed lines at $\pm$55\degree{} to the gray dashed lines at the equator defines the chevron outlining the activity cycle. For reference the BP latitude-time image (Fig.~\pref{f6}C) is shown for Cycle 23. Compare with the zonal flow over four solar cycles (Fig.~\pref{f11}) and for the longer record of sunspot areas (Fig.~\pref{f12}).} \label{f10}
\end{figure}

Using the simple landmarks determined from the observations of cycle 23 above we exploit the Royal Observatory of Belgium and Royal Observatory Greenwich records of sunspot number and area for which the results are shown in Fig.~\pref{f10}. The dashed chevrons start their equatorward motion from $\pm$55\degree{} at times determined using the hemispheric (hSSN) maxima marked by red (north) and blue (south) dashed vertical lines. Deducing from Fig.~\pref{f7} that the low latitude chevrons terminate when the measured sunspot areas exceed 100 millionths we draw the gray dashed vertical lines. 

\begin{figure}[!h]
\epsscale{1.00}
\plotone{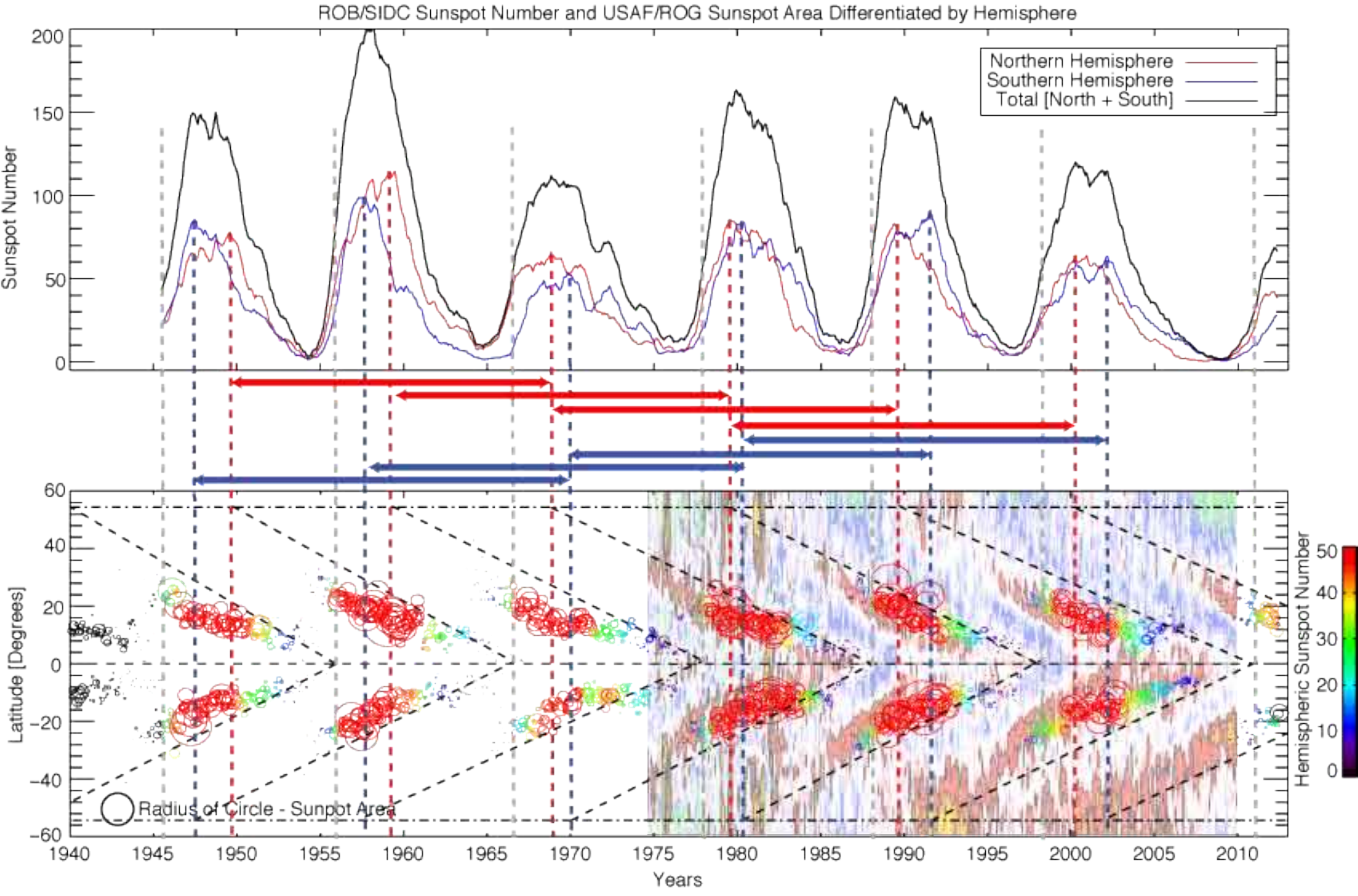}
\caption{Compare with Fig.~\pref{f10}. In this instance we have replaced the BP latitude-time distribution with the zonal flow \-- corresponding to the torsional oscillation \citep[Fig.~\pref{f6}D;][]{2010ApJ...725..658U} \-- constructed from Doppler measurements of the solar photosphere taken at the 150ft Tower at the Mount Wilson Solar Observatory. Note the strong correspondence of the patterns observed with the chevrons drawn using only the Hemispheric Sunspot Number (hSSN) as a guide. Also, note the apparent regularity of the high-latitude starting points of consecutive chevrons.} \label{f11}
\end{figure}

\begin{figure}[!h]
\epsscale{1.00}
\plotone{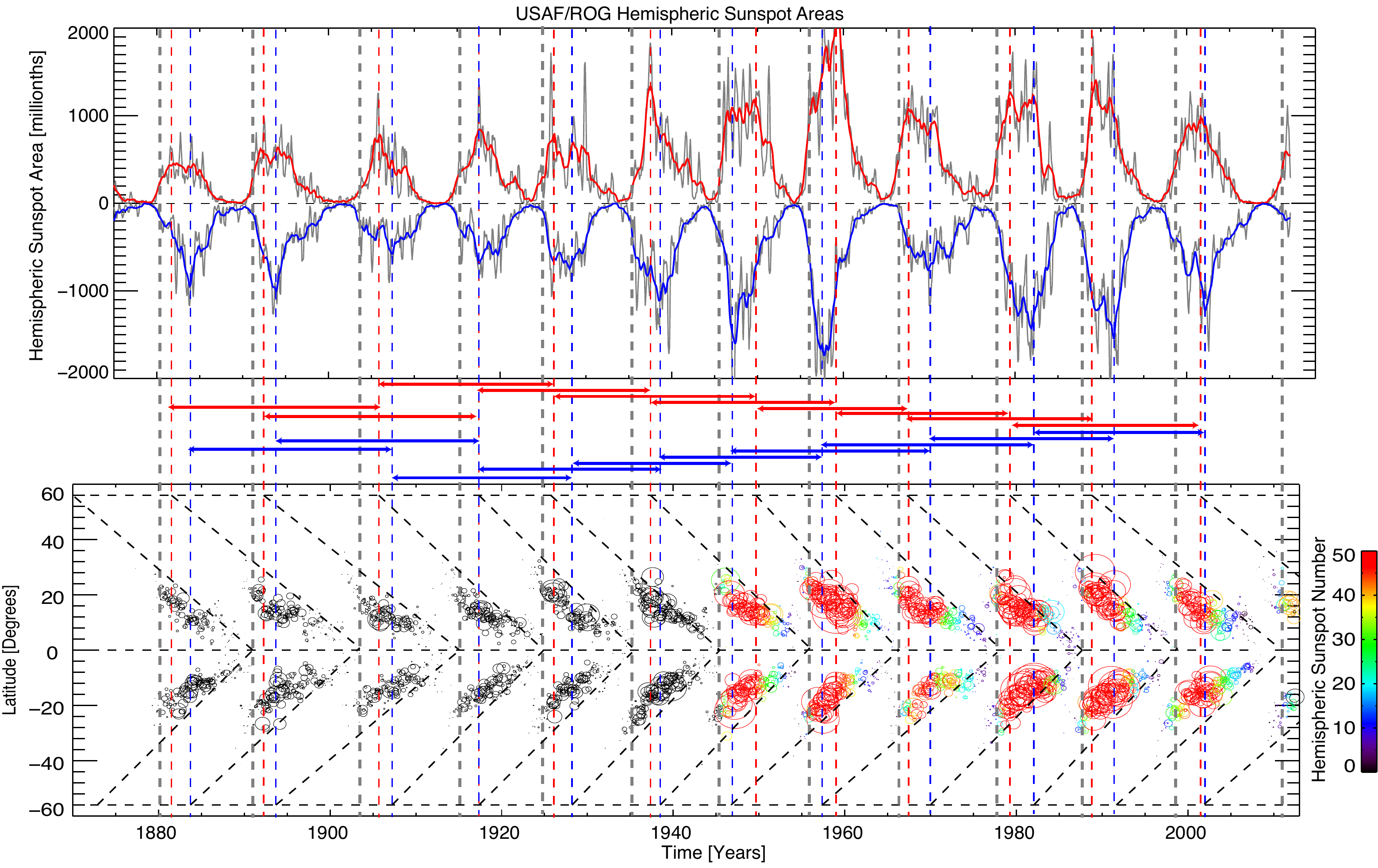}
\caption{Using the USAF/Royal Observatory Greenwich sunspot area distributions as a proxy of hemispheric activity to extend Figs.~\pref{f11} and~\pref{f12} into the nineteenth century. As above the radius of the circles drawn in the lower panel indicate the percentage of the solar disk covered by sunspots (in millionths). Where we have sunspot numbers differentiated by hemisphere (after 1945) we color the circles by their hSSN. In this case we use the hemispheric peak (red \-- north; blue \-- south) in sunspot area to identify the activity maximum in each hemisphere from which the activity band starts at $\pm$55\degree{} latitude and draw the red and blue vertical dashed lines. In this case the grey vertical lines are drawn when the sunspot are exceeds 100 millionths to indicate that the equatorial bands have cancelled/terminated. Again, note the strong consistent 22-year repetition of the high-latitude start of the chevrons in each hemisphere.} \label{f12}
\end{figure}

These simply defined activity band chevrons enclose almost every sunspot in the record (see also Figs.~\pref{f11} and~\pref{f12}). The chevrons of the same sign start very close to 22 years apart (thick colored arrows that are colored to indicate hemisphere). To illustrate the correspondence between these figures and Fig.~\pref{f6} we show the BP density record (Figs.~\pref{f10}) and the zonal flow patterns inferred from Doppler measurements of the solar photosphere taken at the 150ft Tower at the Mount Wilson Solar Observatory \citep[][]{2010ApJ...725..658U}. In the latter example, we see the clear relationship between the pattern flow patterns and the high-low latitude evolution characteristic of the torsional oscillation, of which this is an analog, over the past three decades.

Using {\em only} the sunspot area record and the hemispheric maxima detected therein we can compare consecutive maxima of the same sign somewhat quantitatively in Table~\pref{tbl:one} and Fig.~\pref{f12} where the fits determine a mean evolutionary time of $20.38\pm1.06$ (North) and $21.63\pm{1.52}$ (South) years (and $20.84\pm{1.56}$ years for the average). Also from the table we see that the equatorial bands are not regular in their timing. {\em The transit time from high to low latitude of the activity bands appears to vary inversely with the strength of the cycle \-- magnetically stronger cycles have fast transits while weaker cycles take longer implying that the circulatory speed of cells in the (coupled) system responsible for the activity cycle are critically dependent on the amount of magnetic flux that is loaded onto them.}

\begin{figure}[!h]
\epsscale{0.5}
\plotone{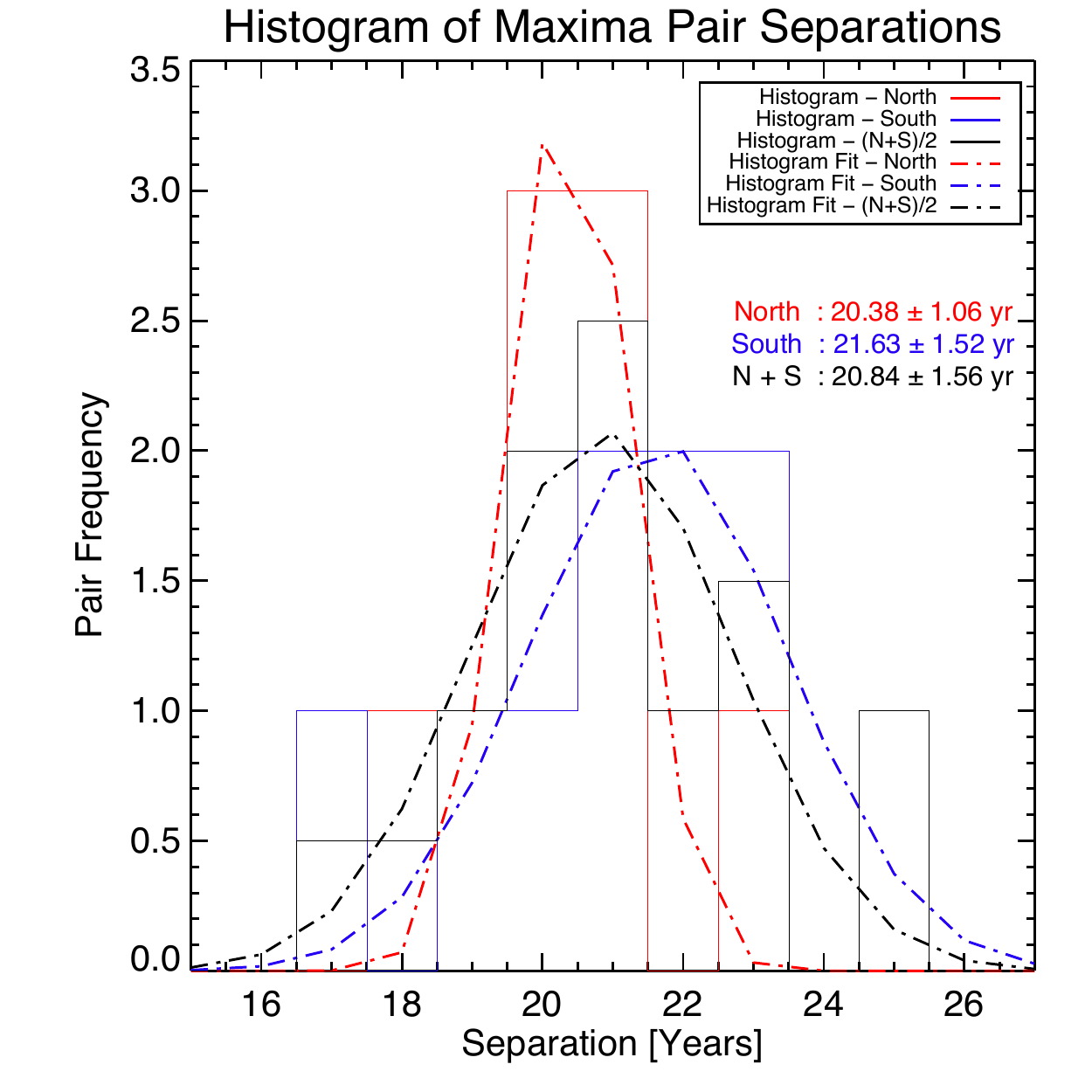}
\caption{The histograms of same-sign solar maxima separation from the hemispheric (North \-- red; South \-- blue) maxima of the USAF/Royal Observatory Greenwich sunspot area distributions shown in Fig.~\pref{f12}. The Gaussian fit to each distribution (dot\--dashed lines) allows us to estimate the mean cycle time of the northern and southern hemispheres.} \label{f13}
\end{figure}

\clearpage

\begin{figure}[!h]
\epsscale{1.00}
\plotone{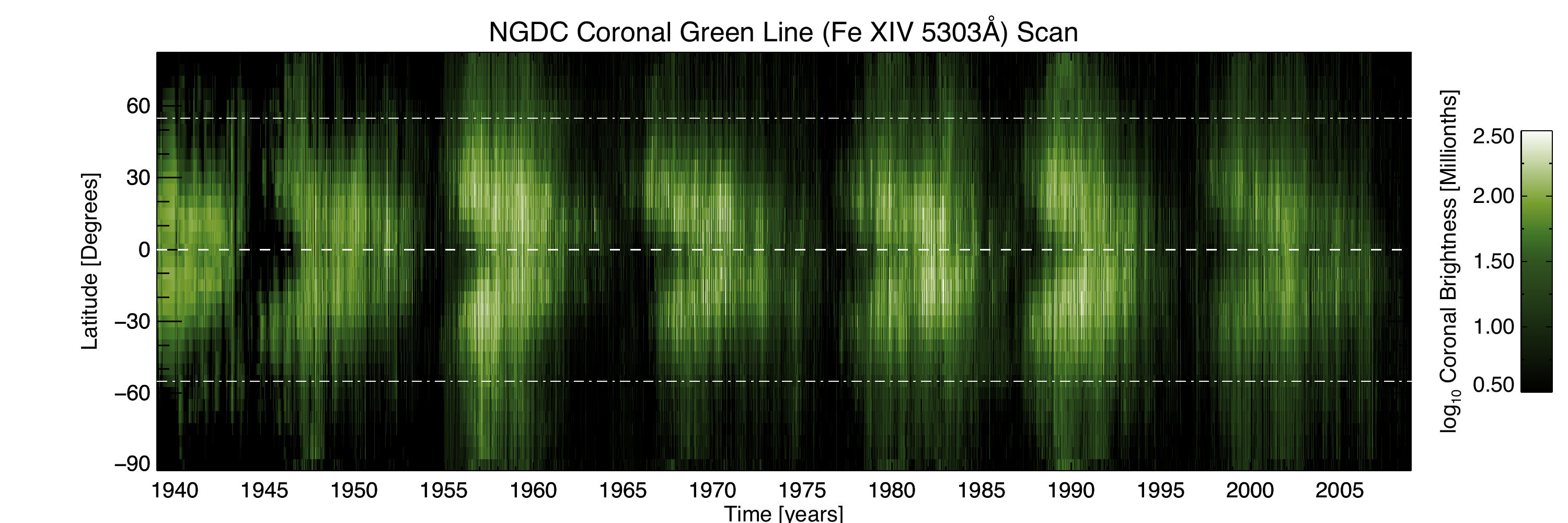}
\caption{The coronal green line (\ion{Fe}{14} 5303\AA) scan from the NGDC. An analog of the data presented in Fig.~\pref{f8} spanning a much longer timescale, the record extends back to 1939. The coronal structure was scanned around the solar limb and integrated in 5\degree{} steps. Like Fig.~\pref{f8} these annular scans show the critical nature of 55\degree{} with respect to coronal structure. We can see that this last solar cycle is not anomalous in restricting coronal emission below 55\degree{} except during the period when the polar reversal is taking place. The equator is shown as a white dashed line while the dot-dashed lines show 55\degree{} in each hemisphere.} \label{f14}
\end{figure}

\subsection{The NGDC Coronal Green Line Record}
The most striking feature in Fig.~\pref{f8} is the dearth of coronal emission above 55\degree{} for the majority of solar cycle 23. Indeed, that latitude appears to be a relatively rigid upper bound for low latitude coronal emission and hence a lower bound for the polar coronal hole in addition to being an upper bound for low-latitude holes (Fig.~\pref{f9}). The only time when there is significant coronal emission at high latitudes is during the progression of oppositely signed magnetic flux that reverses the polarity of the polar region. We also noted above that the duration of the coronal emission surge towards the poles in each hemisphere appears to be consistent with the length of time that the activity band (visualized in the g-node bands in the upper and lower panels) spends at 55\degree{} in that hemisphere before migrating to the equator. To illustrate that these properties of coronal emission with latitude and time are not limited to sunspot cycle 23 consider Fig.~\pref{f14} which shows the NOAA National Geophysical Data Center (NGDC) coronal green line scan archive spanning back to 1939. Again, we see that the bright (hot) coronal emission is largely restricted to latitudes below 55\degree{} with significant excursions poleward only taking place once every $\sim$11 years. This record provides further support to the notion that 55\degree{} is a critical latitude in solar cycle evolution. Indeed, with further work it is possible that observational records like the NGDC green line scan may be used to provide valuable information about the length of time spent by the high latitude bands prior to their migrating equatorward. 

\section{Discussion}\label{discuss}
The length scale on which BPs form is consistent with that of giant cell convection. We interpret the g-nodes, around which those BPs tend to occur \citep[][]{McIntosh2014}, to be the radial component of the deep-seated toroidal magnetic flux band that is rooted at (or near) the tachocline. Additional evidence in support of this interpretation can be found in the rotation rates inferred from the tracking of BPs \citep[e.g.,][]{1978ApJ...219L..55G} and will be explored with contemporary data in a subsequent publication (in preparation). 

\begin{figure}[!h]
\epsscale{0.5}
\plotone{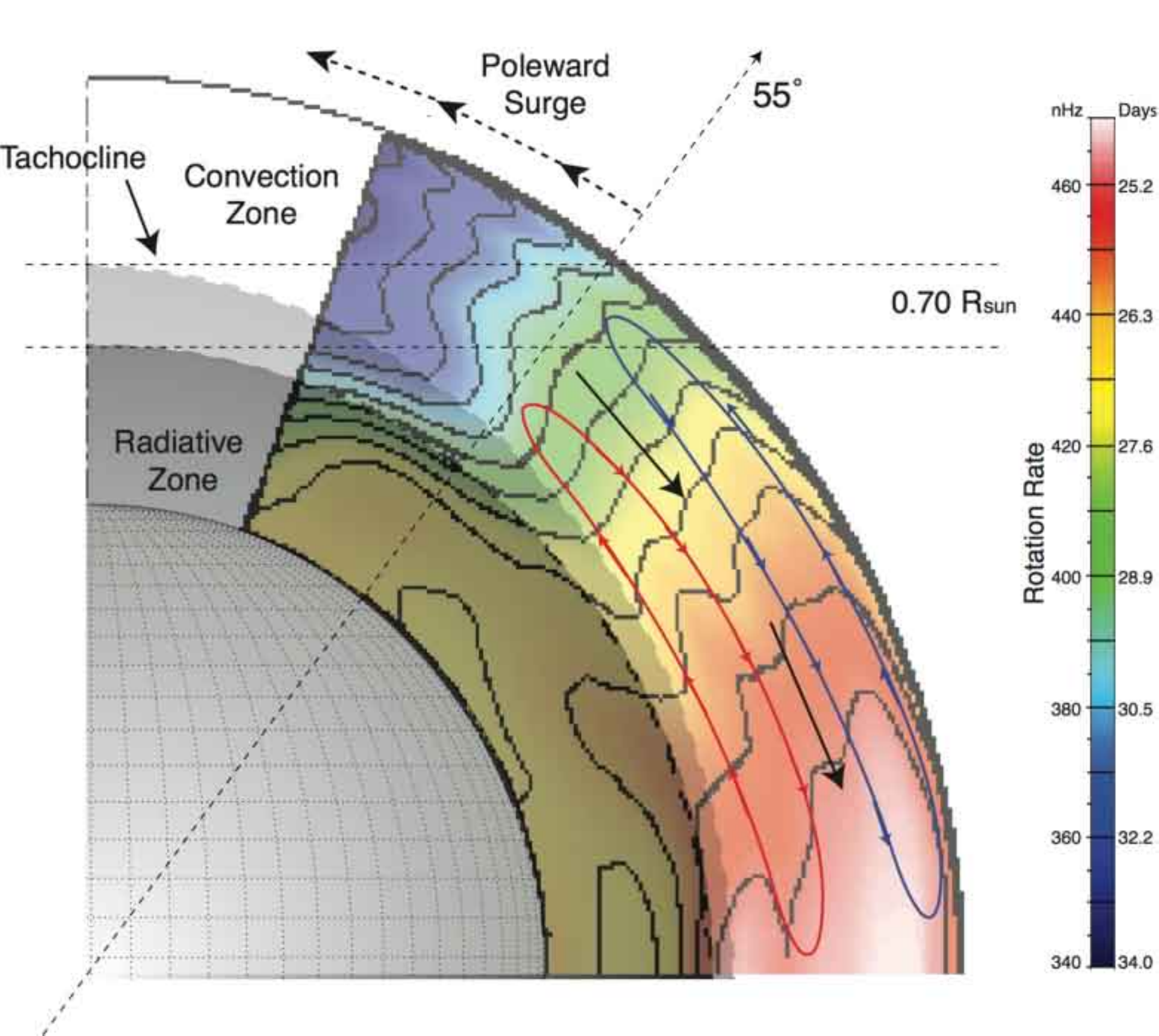}
\caption{A cut of through one quadrant of the Sun to illustrate the possible flow patterns governing the magnetic flux progression. The observations are delineated by the regions above and below 55\degree{} latitude \citep[also seen in the pattern of interior differential rotation][]{1996Sci...272.1300T} as the projected latitude of the depth of the high-latitude radiative zone (darker sphere). We also sketch the region called the tachocline (lighter sphere) and one possible equatorial circulatory pattern below 55\degree{} \citep[][]{2013ApJ...774L..29Z} where we only have a sense of the deep progression of the magnetic elements, and a polar circulation cell above 55\degree{} where we only have a sense of their surface progression.} \label{f16}
\end{figure}

Giant convective cells are aligned with the Sun's rotation axis and so the strong association of many of the phenomena presented above with $\pm$55\degree, as the latitudinal projection of the driving radiative interior and the ``tangent cylinder'' are natural \citep[e.g.,][and Fig.~\pref{f16}]{2005LRSP....2....1M}. In general, BPs (and the magnetic elements associated with the g-nodes) are overlooked in standard latitude-time diagrams due to their relatively small spatial dimension. Only when such features are tracked does their progression become an indicator of coherent magnetic flux emergence preceding sunspot formation \citep[][]{1988Natur.333..748W}. 

While numerical modeling of this system is absolutely required, we speculate that the deep-rooted magnetic flux tube has a simple (and continuous) ability to breach the photosphere \citep[e.g.,][]{2012SoPh..tmp..208W}. The persistent eruption of flux connected to the activity band driven by the large-scale convection above \citep[e.g.,][]{2011ApJ...739L..38N, 2013ApJ...762...73N, 2013SoPh..tmp...20N} permits the progression of the tube from birth to termination to be captured in the surface velocity field, g-node pattern, and associated BPs. The technique of mapping the progression of the underlying magnetic flux tube presented in this paper has a distinct advantage over {\em global} helioseismology methods in inferring the progression of the circulatory pattern during times of profound hemispheric asymmetry.

We have inferred also that the region at latitudes higher than 55\degree{} is important for the progression of the solar cycle. From the approximately constant mean overturn time of the polar cells we propose that they act as clocks for the coupled system. The perturbations to the flow induced by the magnetic field at these high latitudes are small because the average fields themselves are small \citep[$\sim$1G,][]{1958Sci...127.1058B,2013ApJ...763...23S} when compared to that of the equatorial region. Indeed, this evolution of the polar cells can possibly help understand the long-established observational relationship between the polar field strength and the strength of the following sunspot cycle \citep[][]{2010LRSP....7....1H} \-- if less flux erupts, then less is caught in the polar cell, and so less is available to form the flux tube of the cycle 22 years later and so on. Further, when less flux is available at high latitudes the equatorial chevrons migrate slower towards the equator and this process appears to feed back on itself, although in this picture, there is little effect on the interweaved cycle which may be weaker or stronger. This apparent alternating in strength of the cycles could provide some clarity on another observational phenomenon, the Gnevyshev-\"{O}hl or ``odd-even'' alternating progression of sunspot numbers from cycle to cycle \citep[][]{Gnevyshev1948}.

The maps of the torsional oscillation (Fig.~\pref{f6}D, or the zonal flow in Fig.~\pref{f14}) provide a mapping to the surface magnetism discussed above. However, our article is not the first to draw the connection, or pose the questions \citep[][]{1988Natur.333..748W, 2008ASPC..383..335A}: what is the torsional oscillation and what is its connection to the 22-year magnetic cycle? The analysis presented appears to offer some insight into the modulation of the solar cycle. Indeed, it remarkably consistent with the concept of ``The Extended Solar Cycle'' \citep[ESC,][]{1988Natur.333..748W}. The ESC is an observationally derived paradigm which, over the last several decades, has persistently demonstrated an observational connection between helioseismology \citep[][]{2008ASPC..383..335A}, surface magnetic phenomena, and extended coronal features \citep[][]{2013ApJ...765..146M,2013SoPh..282..249T} which display a cyclic behavior with a period approaching the complete 22 years of the magnetic activity cycle. 

Further work is required to understand fully the precise balance between the magnetic field and the impact on the circulatory system, how the magnetic field forms, and how it is loaded into the circulatory system. However, we can see that stronger cycles are faster and weaker cycles are slower and that the strength of the cycle is dependent on the amount of flux advected into the polar cell in the preceding cycle of the same sign. There appears to be a hysteresis in the feedback between the flow and the magnetic field of the equatorial region that can establish rising or falling strength of magnetic activity over longer time scales. Indeed, further work is also required to combine these \soho{} and \sdo{} observations with those from Kitt-Peak, Mount Wilson, and older space records of small-scale magnetic evolution to consistently probe multiple activity cycles.

Unfortunately, many of the solar dynamo models currently employed in the community are impacted by the results presented herein \citep[e.g.,][]{1955ApJ...121..349B,1969ApJ...156....1L,1999ApJ...518..508D,2011Natur.471...80N}. In general, these models require the formation of a new poloidal flux system at the end of every 11-year cycle that can seed the toroidal field of the next 11 years. However, the observations provided above would appear to demonstrate that these activity bands form at very high latitudes, and are overlapping temporally and interacting with each other in the solar interior for more than half of the 22-year period. Therefore, we must explore new ways to explain the complexity of this coupled feedback system. The next generation of self-consistent magnetohydrodynamic models of the solar interior may offer significant clues \citep[][]{2010LRSP....7....3C,2011ApJ...735...46R} as they seem to reproduce a broader range of the observed phenomena. {\em However, if our determination that the regularity of high-latitude evolution/circulation is critical to the modulation of the coupled rotating (magneto-)hydrodynamic system then detailed simulations and careful observations of the general circulation (and especially the polar regions) should reveal the controlling physical processes responsible for the production of the magnetic modulation and its evolution.} 

Motivated by our observational analysis we propose that it is the polar regions, rather than the lower latitudes, that regulate the time scale of the solar activity cycle while the lower latitudes respond. The zonal flow oscillation at high latitudes may set this timescale by hydrodynamic means such as coupling to the radiative interior or it may arise through the oscillating fields of a cyclic dynamo operating in the polar regions. Such a ``polar dynamo'' would account for the distinct evolution of the magnetic signatures above and below $\pm$55\degree{} evident in many of the figures presented herein.


If distinct dynamos do indeed operate at the north and south poles, injecting magnetic flux into lower latitudes at different times then what gives rise to the remarkable synchronization of the lower-latitude activity branches such that the north and south branches of the chevrons terminate at the same point? One possibility might be a nonlinear relaxation of the magnetic field roughly analogous to the ``annealing'' of poloidal flux discussed by \cite{2011sswh.book...39S}. If the polar dynamos inject helical flux systems into lower latitudes as illustrated in Fig.~\pref{f15} and~\pref{f16}  then the merging and cancellation of oppositely-signed toroidal bands in the north and south would produce a coalescence of poloidal fields that may help synchronize the hemispheres. Furthermore, if some of the resulting poloidal field were to thread through the photosphere, this may account for the proliferation of BPs and g-nodes near the tips of the chevrons evident in Figs.~\pref{f6}. Alternatively, injection of flux from the polar dynamo branches may excite a symmetric dynamo wave at lower latitudes. In any case, {\em we propose that high-latitude dynamo action acts as a ``clock'' to set the pace of the solar cycle}.

Naturally, the ``grand challenges'' for any piece of research (observational or otherwise) that claims to understand the modulation of the sunspot cycle are how does the Sun enter {\em and recover} from grand minima, and what can we infer about the next solar cycle (or beyond)? Further, what do the intra\-- and extra\--hemispheric coupling of the activity bands at different phases of the 22-year period mean for flux emergence processes, flare and coronal mass ejections? These matters will be discussed in forthcoming papers in the series. 

To close on a similarly broad canvas, in the study of stellar activity cycles \citep[e.g.,][]{2007ApJ...657..486B} it is time to acknowledge that the 22-year magnetic cycle is the Sun's ``fundamental'' mode and not 11\--years as reflected in the pattern produced by the interaction of the activity bands. It remains to be seen how the star's rotation rate, convection zone depth and other (relatively invariant) properties can combine to produce the spectrum of stellar activity cycles observed, but that are likely mis-characterized. For example, consider how many activity bands per hemisphere would a star that was identical in mass and convection zone depth to the Sun have if it rotated three or five times faster? The Calcium index of that hypothetical star would reflect the ``beat'' of the overlapping bands and not the fundamental. The question must then become how do we identify the fundamental evolutionary timescale  of the surface flux production {\em without} resolved observations? 

\subsection{Summary of Findings}\label{summary}
The findings presented in this paper present a phenomenological explanation for the phases of the cyclic behavior of sunspot numbers and appearance:
\begin{itemize}\itemsep2pt
	\item{The $\sim$11-year sunspot cycle is a result of interaction between the (temporally) overlapping toroidal activity bands of the 22-year magnetic activity cycle. The (oppositely signed; in each hemisphere) activity bands take $\sim$19 years from their emergence at high latitudes ($\pm$55\degree) to reach the equator.}
	\begin{itemize}
		\item{The time between onset of emergence of activity bands of the same sign at high latitude in the same hemisphere is $\sim$20.84 ($\pm$1.56) years.}
	\end{itemize}
	\item{The magnetic activity bands are rooted in the deep interior and small-scale magnetic flux emergence from them allows their passage to be tracked in a set of synoptic observations.}
	\begin{itemize}
		\item{The ``Magnetic Range of Influence'' (MRoI) presents a possible observational link to giant cell convection through the $\sim$100Mm length scales observed.}
		\item{Coronal BrightPoints (BPs) form almost exclusively around the vertices of these giant convective cells, ``g-nodes''.}
		\item{Identifying MRoI g-nodes and BPs allow us to track the progress of the activity bands.}
		\item{The overlapping activity bands match the progression of the torsional oscillation - linking surface magnetism to helioseismic inference.}
		\item{The progression of low-latitude coronal holes also allows us to track the progress of the activity bands.}
		\item{There is only one surge of magnetic activity that is swept into the polar region by the surface meridional circulation. The duration of that surge appears to match the time length of time spent by the activity bands at high latitude prior to starting their migration equatorward.}
	\end{itemize}	
	\item{The intra-hemispheric and extra-hemispheric communication of the activity bands modulates the appearance of sunspots (cf. Fig.~\pref{f8}).}
	\begin{itemize}
		\item{The activity bands form ``Chevrons.'' The overlap in time of these chevrons, when compared directly to the variation of the hemispheric sunspot number, provide the observational clues to understanding the phases of the solar cycle:}
		\begin{itemize}
			\item{When no oppositely activity band exists in the same hemisphere sunspots can abundantly emerge rooted in the existing activity band. There is only only communication across the equator between the activity bands. This defines the rapid sunspot number growth described as the ``Ascending Phase''.}
			\item{The emergence of the new (oppositely signed) high latitude activity band marks a downturn in the occurrence of sunspots in each hemisphere. The two bands mutually interfere, hence reducing the ability of the low-latitude band to produce sunspots. The point of downturn defines ``Solar Maximum'', and marks the start of the ``Declining Phase''.}
			\item{The activity bands are both migrating to the equator under the action of the global circulation that is (probably) a result of the Sun's rotation. The latitudinal offset between the oppositely signed activity bands {\em may} be important for (large) flare production if oppositely sensed flux emergence readily takes place neighboring, or in, a pre-existing active region.}
			\item{When the two low latitude activity bands eventually ``terminate'' at the equator sunspots appear shortly thereafter following the path of the higher latitude bands (now at around 30\degree{} latitude) within only one or two solar rotations. This marks the start of the {\em next} ascending phase.}
			\item{This concept of overlapping, oppositely signed, activity bands significantly changes the paradigm of ``Solar Minimum'', in the context presented solar minimum is a time of maximal overlap and cancellation between the activity bands in their own hemisphere {\em and} across the equator.} 
		\end{itemize}
	\end{itemize}
	\item{Based on the observations studied, the degree of overlap between the chevrons in the equatorial region of the ``extended solar cycle'' dictate the shape, length, and magnitude of the sunspot cycle. These factors likely impact the pattern of solar energetic output \citep[e.g.,][]{2013ApJ...765..146M}. There appears to be a significant feedback between the magnetic field and the circulatory flow:}
	\begin{itemize}
		\item{More magnetic flux in the equatorial activity bands speed up the transit from high to low latitude and give rise to strong cycles}
		\item{Less magnetic flux in the equatorial activity bands slows down the progression and lengthens the overlap of the chevrons. This increasing overlap results in a weak cycle and less sunspots.}
		\item{The latter appears to the be a ever-decreasing state \-- the reduction of flux in the equatorial region begets another.}
		\item{The transit time may be an robust indicator of the strength of the subsequent sunspot cycle.}
	\end{itemize}
\end{itemize}

\begin{figure}[!h] 
\epsscale{1.0}
\plotone{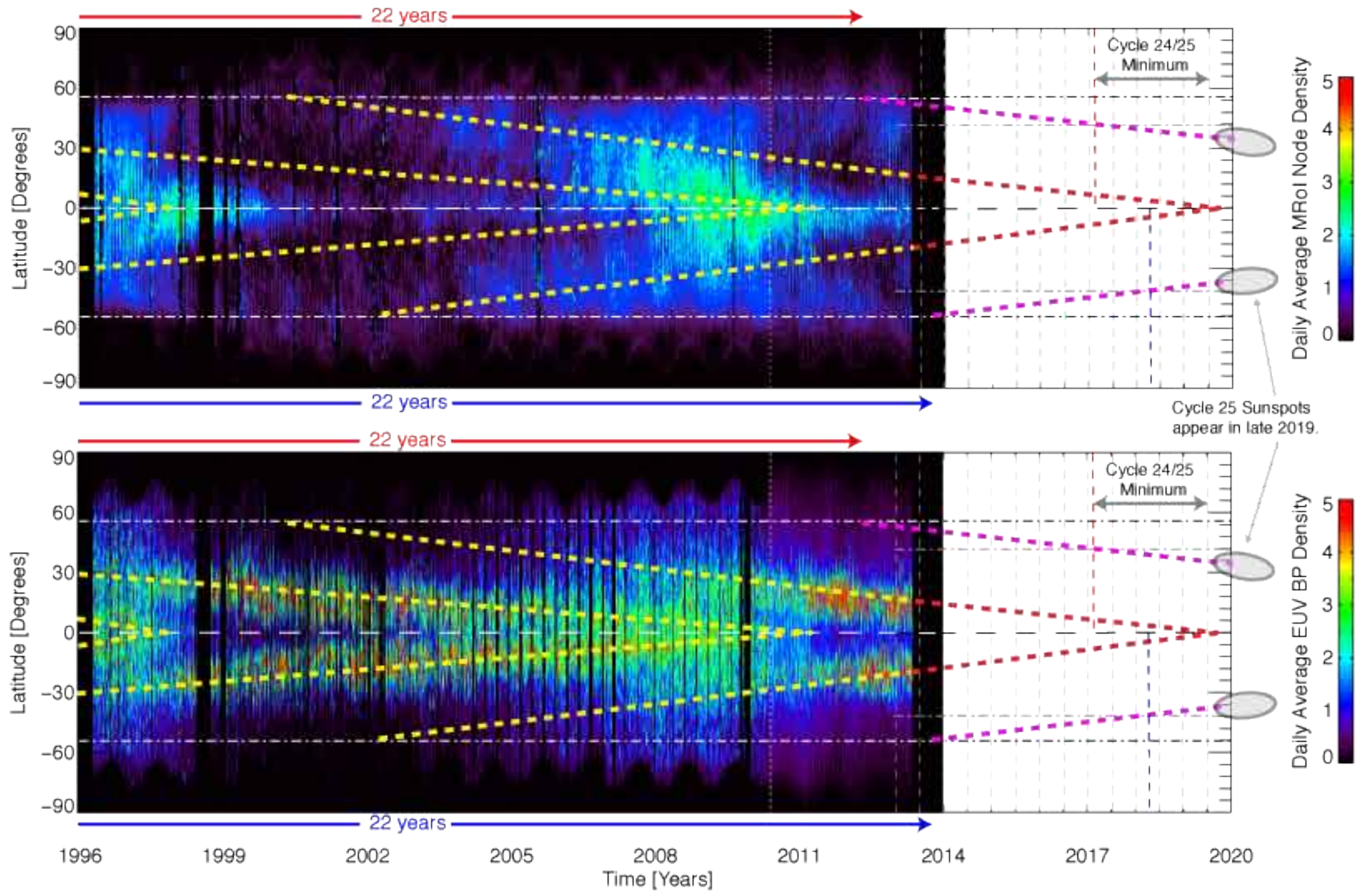}
\caption{Using the panels of Fig.~\pref{f6} to extrapolate the termination of the BP chevrons based on the metrics determined above. The chevrons (yellow dashed lines) are taken from Fig.~\pref{f6} and the faint vertical dashed gray lines are 6 months apart. The red-dashed lines are the linear continuation of the yellow cycle 24 bands. The cycle 25 bands appear as pink dashed lines starting at high latitudes 22 years from the start of their cycle 23 counterparts (derived from Figs.~\pref{f10} and ~\pref{f11}). The vertical red and blue dashed lines indicate possible onset of solar minimum conditions in the northern and southern hemispheres.} \label{f17}
\end{figure}

\section{Forecasting Sunspot Cycle 25 Onset}\label{forecast}
The gradient of the fitted straight lines, and their associated error, can be used to project the migration of the activity bands to a point when they will cross (as shown in Fig. 7). The crossing point indicates the time at which we expect the termination of the cycle 24 bands and the strong emergence of cycle 25 sunspots may occur. This ``forecast'' is made under the assumption that the activity band migration continues at the same linear rate, a assumption that will need to be relaxed in the future once the entire migratory period has been monitored. The pink dashed lines shown in Fig.~\pref{f17} assume that the gradient of latitudinal migration towards the equator does not vary from the fitted value while the dotted lines reflect the minimum and maximum permissible gradients to the fit within the linear assumption. We see that the inner dotted lines cross in the second half of 2017 while the outer dotted lines cross in the last quarter of 2019, the region between these points has been shaded. If the activity bands slow to the values displayed by the cycle 22 and cycle 23 in their last few years then lines shown here will become earliest dates for the cycle 24 termination.

\section{Conclusion}\label{conclude}
We conclude that observations covering the past seventeen years with \soho{} and \sdo{} indicate that the spatio-temporal evolution of the 11-year sunspot cycle is a result of interaction between the temporally overlapping activity bands belonging to the 22-year magnetic activity cycle. Synoptic monitoring of the solar photosphere and corona will continue in order to test this hypothesis and validate our forecast that the sunspots of solar cycle 25 will begin to appear as early as the of this decade.

\acknowledgements
The data used in this paper are openly available from the {\em SOHO}, {\em SDO}, and the Virtual Solar Observatory (VSO; \url{http://virtualsolar.org}) data archives. SWM, RJL, and ARD were partly funded by NASA grants (NNX08AU30G, NNX08AL23G, NNM07AA01C \-- {\em Hinode}, NNG09FA40C \-- {\em IRIS}). AVM is supported by NASA grant NNG04EA00C ({\em SDO}/AIA). {\em SOHO} is a project of international collaboration between ESA and NASA. NCAR is sponsored by the National Science Foundation. SWM is grateful to Dick Altrock, Sara Martin, Roger Ulrich, Mark Miesch, Matthias Rempel, Yuhong Fan, Eugene Parker for helpful discussions. NCAR is sponsored by the National Science Foundation.).

\clearpage

\begin{table}
\caption{This table contains the times of consecutive (same\--signed) maxima, their difference ($\delta$; years) and high-low latitude transit time ($\tau$; years) for each hemisphere determined from Fig.~\pref{f12}.}\label{tbl:one}
    \begin{tabular}{ccccccc}
    Cycle Pair & Maxima (N) & Maxima (S)  & $\delta_{N}$ & $\delta_{S}$ & $\tau_{N}$ & $\tau_{S}$  \\ \hline
    14 \-- 12  & 1905.75, 1884.00 & 1907.08, 1883.83 & 21.75 & 23.25 &  19.25 & 17.92 \\
    15 \-- 13  & 1917.58, 1892.50 & 1919.50, 1893.58 & 25.08 & 25.92 &  19.92 & 18.00 \\
    16 \-- 14  & 1925.92, 1905.75 & 1926.08, 1907.08 & 20.17 & 19.00 &  19.18 & 19.02 \\
    17 \-- 15  & 1937.50, 1917.58 & 1939.67, 1919.50 & 19.92 & 20.17 &  20.50 & 18.33 \\
    18 \-- 16  & 1949.17, 1925.92 & 1947.17, 1926.08 & 23.25 & 21.08 &  19.03 & 21.03 \\
    19 \-- 17  & 1959.08, 1937.50 & 1956.83, 1939.67 & 21.58 & 17.17 &  19.42 & 21.67 \\
    20 \-- 18  & 1968.00, 1949.17 & 1970.08, 1947.17 & 18.83 & 22.92 &  19.50 & 17.42 \\
    21 \-- 19  & 1979.67, 1959.08 & 1980.33, 1956.83 & 20.58 & 23.50 &  19.33 & 18.67 \\
    22 \-- 20  & 1989.08, 1968.00 & 1991.08, 1970.08 & 21.08 & 21.00 &  21.92 & 19.92 \\
    23 \-- 21  & 2000.50, 1979.67 & 2002.58, 1980.33 & 20.83 & 22.25 &  \-- & \-- \\ \hline
    \end{tabular}
\end{table}


\begin{thebibliography}{}
\bibitem[Altrock et al.(2008)]{2008ASPC..383..335A} Altrock, R., Howe, R., \& Ulrich, R.\ 2008, Subsurface and Atmospheric Influences on Solar Activity, 383, 335 
\bibitem[Babcock \& Babcock(1955)]{1955ApJ...121..349B} Babcock, H.~W., \& Babcock, H.~D.\ 1955, \apj, 121, 349
\bibitem[Babcock(1959)]{1959ApJ...130..364B} Babcock, H.~D.\ 1959, \apj, 130, 364
\bibitem[Babcock \& Livingston(1958)]{1958Sci...127.1058B} Babcock, H.~D., \& Livingston, W.~C.\ 1958, Science, 127, 1058
\bibitem[B{\"o}hm-Vitense(2007)]{2007ApJ...657..486B} B{\"o}hm-Vitense, E.\ 2007, \apj, 657, 486 
\bibitem[Brun et al.(2004)]{2004ApJ...614.1073B} Brun, A.~S., Miesch, M.~S., \& Toomre, J.\ 2004, \apj, 614, 1073 
\bibitem[Charbonneau(2010)]{2010LRSP....7....3C} Charbonneau, P.\ 2010, Living Reviews in Solar Physics, 7, 3 
\bibitem[Charbonneau, Dikpati, \& Gilman(1999)]{1999ApJ...526..523C} Charbonneau, P., Dikpati, M., \& Gilman, P.~A.\ 1999, \apj, 526, 523 
\bibitem[Delaboudini{\`e}re et al.(1995)]{1995SoPh..162..291D} Delaboudini{\`e}re, J.-P., et al.\ 1995, \solphys, 162, 291
\bibitem[Dikpati \& Charbonneau(1999)]{1999ApJ...518..508D} Dikpati, M., \& Charbonneau, P.\ 1999, \apj, 518, 508  
\bibitem[Golub \& Vaiana(1978)]{1978ApJ...219L..55G} Golub, L., \& Vaiana, G.~S.\ 1978, \apjl, 219, L55
\bibitem[Gnevyshev \& \"{O}hl(1948)]{Gnevyshev1948} Gnevyshev, M.~N. and \"{O}hl, A.~I., 1948, Astron. Zh., 25, 18
\bibitem[Hale, Ellerman, Nicholson, \& Joy(1919)]{1919ApJ....49..153H} Hale, G.~E., Ellerman, F., Nicholson, S.~B., \& Joy, A.~H.\ 1919, \apj, 49, 153 
\bibitem[Hale(1924)]{1924Natur.113..105H} Hale, G.~E.\ 1924, \nat, 113, 105 
\bibitem[Hara \& Nakakubo-Morimoto(2003)]{2003ApJ...589.1062H} Hara, H., \& Nakakubo-Morimoto, K.\ 2003, \apj, 589, 1062 
\bibitem[Harvey(1992)]{1992ASPC...27..335H} Harvey, K.~L.\ 1992, Ast. Soc. Pac. Conf. Ser., 27, 335
\bibitem[Hathaway(1996)]{1996ApJ...460.1027H} Hathaway, D.~H.\ 1996, \apj, 460, 1027
\bibitem[Hathaway(2010)]{2010LRSP....7....1H} Hathaway, D.~H.\ 2010, Living Reviews in Solar Physics, 7, 1
\bibitem[Howe et al.(2000)]{2000Sci...287.2456H} Howe, R., Christensen-Dalsgaard, J., Hill, F., Komm, R.~W., Larsen, R.~M., Schou, J., Thompson, M.~J., \& Toomre, J.\ 2000, Science, 287, 2456
\bibitem[Howe(2009)]{2009LRSP....6....1H} Howe, R.\ 2009, Living Reviews in Solar Physics, 6, 1
\bibitem[Krista \& Gallagher(2009)]{2009SoPh..256...87K} Krista, L.~D., \& Gallagher, P.~T.\ 2009, \solphys, 256, 87 
\bibitem[Labonte \& Howard(1982)]{1982SoPh...75..161L} Labonte, B.~J., \& Howard, R.\ 1982, \solphys, 75, 161 
\bibitem[Lemen et al.(2012)]{2012SoPh..275...17L} Lemen, J.~R., et al.\ 2012, \solphys, 275, 17
\bibitem[Leighton(1969)]{1969ApJ...156....1L} Leighton, R.~B.\ 1969, \apj, 156, 1 
\bibitem[Maunder(1904)]{1904MNRAS..64..747M} Maunder, E.~W.\ 1904, \mnras, 64, 747
\bibitem[McIntosh \& Gurman(2005)]{2005SoPh..228..285M} McIntosh, S.~W., \& Gurman, J.~B.\ 2005, \solphys, 228, 285 
\bibitem[McIntosh, Davey, \& Hassler(2006)]{2006ApJ...644L..87M} McIntosh, S.~W., Davey, A.~R., \& Hassler, D.~M.\ 2006, \apjl, 644, L87
\bibitem[McIntosh(2007)]{2007ApJ...670.1401M} McIntosh, S.~W.\ 2007, \apj, 670, 1401
\bibitem[{{McIntosh} {et~al.}(2007)}]{McIntosh2007a} {McIntosh}, S.~W., {et~al.} 2007, \apj, 654, 650
\bibitem[{{McIntosh}, {Leamon} \& {De Pontieu}(2010)}]{McIntosh2010} McIntosh, S.~W., Leamon, R.~J. \& De Pontieu, B.\ 2010, \apj, 727, 7
\bibitem[McIntosh et al.(2013)]{2013ApJ...765..146M} McIntosh, S.~W., Leamon, R.~J., Gurman, J.~B., et al.\ 2013, \apj, 765, 146 
\bibitem[McIntosh et al.(2014a)]{McIntosh2014} McIntosh, S.~W., Wang, X, Leamon, R.~J., \& Scherrer, P.~H., 2014a, \apjl, 748, 32
\bibitem[McIntosh et al.(2014b)]{McIntosh2014b} McIntosh, S.~W., et~al., 2014b, ``Deciphering Solar Magnetic Activity II: On The Quasi-Periodic Forcing of The SunÕs Eruptive, Radiative, and Particulate Output'', Nature Communications (Submitted).
\bibitem[Miesch(2005)]{2005LRSP....2....1M} Miesch, M.~S.\ 2005, Living Reviews in Solar Physics, 2, 1
\bibitem[Nandy, Mu{\~n}oz-Jaramillo, \& Martens(2011)]{2011Natur.471...80N} Nandy, D., Mu{\~n}oz-Jaramillo, A., \& Martens, P.~C.~H.\ 2011, \nat, 471, 80 
\bibitem[Nelson et al.(2011)]{2011ApJ...739L..38N} Nelson, N.~J., Brown, B.~P., Brun, A.~S., Miesch, M.~S., \& Toomre, J.\ 2011, \apjl, 739, L38
\bibitem[Nelson et al.(2014)]{2013SoPh..tmp...20N} Nelson, N.~J., Brown, B.~P., Sacha Brun, A., Miesch, M.~S., \& Toomre, J.\ 2014, \solphys, 289, 441
\bibitem[Nelson et al.(2013)]{2013ApJ...762...73N} Nelson, N.~J., Brown, B.~P., Brun, A.~S., Miesch, M.~S., \& Toomre, J.\ 2013, \apj, 762, 73 
\bibitem[Racine et al.(2011)]{2011ApJ...735...46R} Racine, {\'E}., Charbonneau, P., Ghizaru, M., Bouchat, A., \& Smolarkiewicz, P.~K.\ 2011, \apj, 735, 46 
\bibitem[Rempel(2007)]{2007ApJ...655..651R} Rempel, M.\ 2007, \apj, 655, 651 
\bibitem[Sheeley(2005)]{2005LRSP....2....5S} Sheeley, N.~R., Jr.\ 2005, Living Reviews in Solar Physics, 2, 5 
\bibitem[Schwabe(1844)]{1844AN.....21..233S} Schwabe, M.\ 1844, Astronomische Nachrichten, 21, 233
\bibitem[Scherrer et al.(1995)]{1995SoPh..162..129S} Scherrer, P.~H., et al.\ 1995, \solphys, 162, 129 
\bibitem[Scherrer et al.(2012)]{2012SoPh..275..207S} Scherrer, P.~H., et al.\ 2012, \solphys, 275, 207 
\bibitem[Spruit(2011)]{2011sswh.book...39S} Spruit, H.~C.\ 2011, The Sun, the Solar Wind, and the Heliosphere, 39 
\bibitem[Svalgaard \& Kamide(2013)]{2013ApJ...763...23S} Svalgaard, L., \& Kamide, Y.\ 2013, \apj, 763, 23 
\bibitem[Tappin \& Altrock(2013)]{2013SoPh..282..249T} Tappin, S.~J., \& Altrock, R.~C.\ 2013, \solphys, 282, 249
\bibitem[Thompson et al.(1996)]{1996Sci...272.1300T} Thompson, M.~J., et al.\ 1996, Science, 272, 1300 
\bibitem[Ulrich(2010)]{2010ApJ...725..658U} Ulrich, R.~K.\ 2010, \apj, 725, 658 
\bibitem[Vaiana, Krieger, \& Timothy(1973)]{1973SoPh...32...81V} Vaiana, G.~S., Krieger, A.~S., \& Timothy, A.~F.\ 1973, \solphys, 32, 81
\bibitem[Weber et al.(2013)]{2012SoPh..tmp..208W} Weber, M.~A., Fan, Y., \& Miesch, M.~S.\ 2013, \solphys, 287, 239 
\bibitem[Wilson et al.(1988)]{1988Natur.333..748W} Wilson, P.~R., Altrock, R.~C., Harvey, K.~L., Martin, S.~F., \& Snodgrass, H.~B.\ 1988, \nat, 333, 748 
\bibitem[Zhao et al.(2013)]{2013ApJ...774L..29Z} Zhao, J., Bogart, R.~S., Kosovichev, A.~G., Duvall, T.~L., Jr., \& Hartlep, T.\ 2013, \apjl, 774, L29 






























\end{thebibliography}
\end{document}